\documentclass[aps,prb,twocolumn,showpacs,superscriptaddress,floatfix,10pt]{revtex4-1}

\usepackage{graphicx}
\usepackage{amsmath}
\usepackage{epsfig}
\usepackage{helvet}
\usepackage{amssymb}

\newcommand{\bk}[1]{\mbox{$\langle #1 \rangle$}}

\newcommand{\norm}[1]{\mbox{$\left\| #1 \right\|$}}
\newcommand{\bnorm}[1]{\big\lVert#1\big\rVert}

\def\R{{\mathcal R}}
\def\F{{\mathcal F}}
\def\G{{\mathcal G}}
\def\tF{{\mathcal{\tilde F}}}

\begin{document}

\title{Algorithms for tensor network renormalization}

\author{G. Evenbly}
\affiliation{Department of Physics and Astronomy, University of California, Irvine, CA 92697-4575 USA}
\email{gevenbly@uci.edu}
\date{\today}

\begin{abstract}
We discuss in detail algorithms for implementing tensor network renormalization (TNR) for the study of classical statistical and quantum many-body systems. Firstly, we recall established techniques for how the partition function of a $2D$ classical many-body system or the Euclidean path integral of a $1D$ quantum system can be represented as a network of tensors, before describing how TNR can be implemented to efficiently contract the network via a sequence of coarse-graining transformations. The efficacy of the TNR approach is then benchmarked for the $2D$ classical statistical and $1D$ quantum Ising models; in particular the ability of TNR to maintain a high level of accuracy over sustained coarse-graining transformations, even at a critical point, is demonstrated.
\end{abstract}


\maketitle

\tableofcontents
\
\section{Introduction} \label{sect:Intro}
Tensor network renormalization\cite{TNR} (TNR) is a recently introduced approach for coarse-graining tensor networks, with application to the efficient simulation of classical statistical many-body systems and quantum many-body systems. A key feature of TNR that differentiates it from previous methods for coarse-graining tensor networks, including Levin and Nave's tensor renormalization group\cite{TRG} (TRG) as well as other subsequently developed approaches\cite{TRGplus, TERG, SRG, TEFR, TRGenv, TRG3D, SpinNet, TRG3DB}, is the use of unitary \emph{disentanglers} in TNR that function to allow removal of all short-ranged correlation at each length scale. This proper removal of short-ranged correlation allows TNR to resolve significant computational and conceptual problems encountered by previous methods.

Despite the success and usefulness of TRG, it is known to suffer a computational breakdown when at or near a critical point\cite{TRG}, where the cost of maintaining at accurate effective description of the system grows quickly with coarse-graining step, due to the accumulation of short-ranged correlation. The use of disentanglers allows TNR to prevent this accumulation, such that TNR can maintain an description over repeated coarse-graining steps, or equivalently for very large system sizes, without requiring a growth of computational cost. Previous methods for coarse-graining tensor networks, such as TRG, are also conceptually problematic if they are to be interpreted as generating a renormalization group\cite{RGReview} (RG) flow in the space of tensors, in that they do not reproduce the expected structure of RG fixed points. This flaw was partially resolved with the proposal of tensor entanglement filtering renormalization (TEFR) in Ref.\onlinecite{TEFR}, which reproduces the proper structure of gapped RG fixed points. On the other hand, TNR fully resolves these problems, reproducing the proper structure of gapped fixed points as well as producing scale-invariant fixed points when applied to critical systems corresponding to discrete versions of conformal field theories\cite{CFT1, CFT2} (CFT), thus correctly realizing Wilson’s RG ideas\cite{Wilson} on tensor networks. By capturing scale-invariance, and producing a rescaling transformation for the lattice consistent with conformal transformations of the field theory\cite{TNRCFT}, TNR can produce an accurate description of the fixed-point RG map, from which the critical data characterizing the CFT can then be extracted.

The use disentanglers in TNR, and their success in preventing the retention and accumulation of short-ranged degrees of freedom, is closely related to the use and success of disentanglers in entanglement renormalization\cite{ER} (ER) and in the multi-scale entanglement renormalization ansatz\cite{MERA} (MERA). This connection was formalized in Ref.\onlinecite{TNRtoMERA}, which showed that TNR, when applied to the Euclidean path integral of a quantum Hamiltonian $H$, can generate a MERA for ground, excited and thermal states of $H$. Thus TNR also provides an alternative to previous algorithms\cite{Algorithms} based upon variational energy minimization for obtaining optimized MERA, and also allows methods developed for extracting scale-invariant data from quantum critical systems using MERA\cite{criticality1, criticality2, criticality3, criticality4, criticality5} to be generalized to classical statistical systems.  

In this manuscript we introduce the numeric algorithms required to implement TNR for the study of $2D$ classical or $1D$ quantum many-body systems. Due to the use of disentanglers, which are key to the TNR approach, implementation of TNR requires more sophisticated optimization strategies than have been necessary in previous tensor RG approaches. This manuscript is organized as follows. First we discuss the standard techniques through which the partition function of a classical system or the Euclidean path integral of a quantum system can be expressed as a tensor network. Then we discuss the general principle of \emph{local approximations} on which tensor RG schemes are based, before detailing the particular projective truncations involved in the TNR approach. Optimization algorithms for the implementation of TNR are then presented, and their performance benchmarked in in the $2D$ classical and $1D$ quantum Ising models.

\section{Tensor network representations of many-body systems} \label{sect:TN}
In this section we discuss approaches for which the partition function of a $2D$ classical statistical system or the Euclidean path integral of a $1D$ quantum system can each be expressed as a square lattice network, which is the starting point for the TNR approach [and other tensor renormalization methods, such as TRG, in general].

\subsection{Classical many-body systems} \label{sect:TNclassical}
\begin{figure}[!t!b]
\begin{center}
\includegraphics[width=8cm]{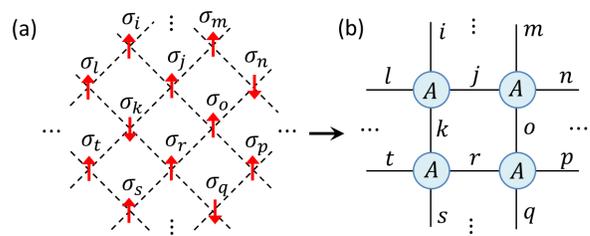}
\caption{(a) A square lattice of classical spins $\sigma \in \{+1,-1\}$. (b) The partition function of the classical system can be encoded as a square network (tilted $45^\circ$ with respect to the spin lattice) of four-index tensors $A_{ijkl}$, with a tensor sitting in the center of every second plaquette of spins. Here each tensor $A$ encodes the Boltzmann weights of associated to the interactions of spins on the edges of the plaquette, see Eq.\ref{eq:A}.}
\label{fig:Partition}
\end{center}
\end{figure}

Here we describe an approach for expressing the partition function $Z$ at temperature $T$ of a $2D$ classical statistical system, 
\begin{equation} \label{eq:Partition}
Z = \sum_{\{\sigma\}} e^{-H(\{\sigma\})/T},
\end{equation}
 as a network of tensors. As a concrete example let us the classical Ising model on the square lattice, with Hamiltonian functional,
\begin{equation} \label{eq:HamFunct}
H\left( \{ \sigma \} \right) =  - \sum_{\left\langle i,j \right\rangle}  \sigma _i\sigma _j,
\end{equation}
where $\sigma_i \in \{+1,-1\}$ is an Ising spin on site $i$. We construct a representation of the partition function as a square-lattice tensor network composed of copies of a four-index tensor $A_{ijkl}$, where a tensor sits in the center of every second plaquette of Ising spins according to a checkerboard tiling, such that the square lattice of $A$ tensors is tilted $45^\circ$ with respect to the lattice of Ising spins, see also Fig.\ref{fig:Partition}. Notice that this corresponds to having one tensor $A$ for every two spins. We define the tensor $A$ to encode the four Boltzmann weights $e^{\sigma_i\sigma_j/T}$ of the Ising spin interactions on the edges of the plaquette on which it sits,
\begin{equation} \label{eq:A}
A_{ijkl} = e^{\left(\sigma_i\sigma_j + \sigma_j\sigma_k + \sigma_k\sigma_l + \sigma_l\sigma_i \right)/T},
\end{equation}
such that the partition function is then given by the sum over all indices,
\begin{equation} \label{eq:Z}
Z = \sum_{ijk\cdots} A_{ijkl}  A_{mnoj}  A_{krst}  A_{opqr} \cdots. 
\end{equation} 
This construction for expressing the partition function as a tensor network can be employed for any model with nearest-neighbor interactions on the square-lattice, and can also be generalized to other lattice geometries and to models with longer range interactions. 

\subsection{Quantum many-body systems} \label{sect:TNquantum}
\begin{figure}[!t!b]
\begin{center}
\includegraphics[width=8cm]{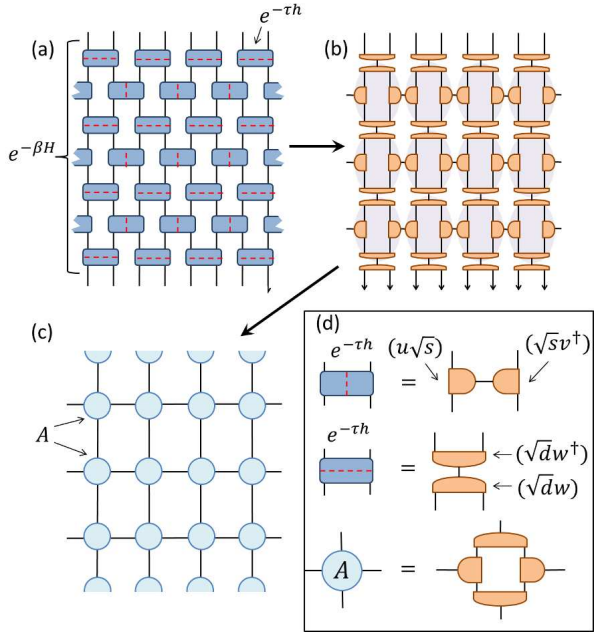}
\caption{(a) The imaginary time-evolution operator $e^{-\beta H}$ is expressed via Suzuki-Trotter expansion as a product of two-site gates $e^{-\tau h}$. Red dashed lines denote how the gates are to be decomposed at the next step. (b) The two-site gates $e^{-\tau h}$ are decomposed into a product of ternary tensors according to either a horizontal partition (accomplished via singular value decomposition) or a vertical partition (accomplished via eigen-decomposition). (c) Groups of four ternary tensors are contracted together to form four-index tensors $A$. (d) Depictions of the vertical and horizontal partitions of the two-site gates $e^{-\tau h}$, and definition of the four index tensors $A$.}
\label{fig:PathIntegral}
\end{center}
\end{figure}

Here we describe how, given a local Hamiltonian $H$ for a $1D$ quantum system, an arbitrarily precise tensor network representation of the Euclidean time evolution operator $e^{-\beta H}$ can be obtained using a Suzuki-Trotter decomposition \cite{Suzuki}. We assume, for simplicity, that Hamiltonian $H$ is a sum of identical nearest-neighbor terms $h$,
\begin{equation} \label{eq:H}
H = \sum_r h_{r,r+1}.
\end{equation}
We begin by expanding the time evolution operator as a product of evolutions over some small time step $\tau$,
\begin{equation} \label{eq:BH}
{e^{ - \beta H}} = {\left( {{e^{ - \tau H}}} \right)^{\left( {\beta /\tau } \right)}}.
\end{equation}
The evolution ${{e^{ - \tau H}}}$ over small time step $\tau$ may then be approximated,
\begin{equation} \label{eq:tH}
{e^{ - \tau H}} \approx {e^{ - \tau {H_{{\text{odd}}}}}}{e^{ - \tau {H_{{\text{even}}}}}}
\end{equation}
where $H_{\text{odd}}$ and $H_{\text{even}}$ represent the contribution to $H$ given from sites $r$ odd or $r$ even respectively, and an error of order $O(\tau)$ has been introduced. [Note that one can obtain an error $O(\tau^n)$, $n>1$, by using a higher order Suzuki-Trotter decomposition\cite{HigherOrder}]. Since $H_{\text{odd}}$ is a sum of terms that act on different sites and therefore commute, $e^{-\tau H_{\text{odd}}}$ is simply a product of two-site gates, and similarly for $e^{-\tau H_{\text{even}}}$,
\begin{align} \label{eq:oeH}
e^{-\tau H_{\text{odd}}} &= \prod_{\text{odd} ~ r} e^{-\tau h_{r,r+1}},\nonumber\\
e^{-\tau H_{\text{even}}} &= \prod_{\text{even} ~ r} e^{-\tau h_{r,r+1}}.
\end{align}
Thus, if one regards each two-site gate $e^{-\tau h}$ as a four index tensor and Eqs.\ref{eq:oeH} and \ref{eq:tH} are substituted into Eq.\ref{eq:BH}, a representation of the Euclidean path integral $e^{-\beta H}$ as a square-lattice tensor network is obtained, see also Fig. \ref{fig:PathIntegral}(a). Note that this representation of ${e^{ - \beta H}}$ has incurred an error of order $O(\beta \tau)$, which can be diminished through use of a smaller time step $\tau$.

While this network could potentially serve as the starting point for the TNR approach [or other algorithm for the renormalization of a tensor network] it is desirable to perform some preliminary manipulations before employing TNR. This initial manipulation involves (i) a transformation that maps to a new square-lattice network tilted $45^\circ$ with respect to the initial network, followed by (ii) coarse-graining in the Euclidean time direction. Given that the initial tensor network is highly anisotropic for small time step $\tau$, as the operator $e^{-\tau h}$ is very close to the identity, step (ii) is useful to obtain a tensor network representation of $e^{-\beta H}$ that is closer to being isotropic [and thus more suitable as a starting point for TNR].

Step (i) is accomplished by performing a modified step of the TRG algorithm as follows. The singular value decomposition (SVD) is taken across a vertical partition of the gate $e^{-\tau h}$,
\begin{equation} \label{eq:vert}
{e^{ - \tau h}} = \left( {u\sqrt s } \right)\left( {\sqrt s {v^\dag }} \right),
\end{equation}
where the root of the singular weights $s$ has been absorbed into each of the unitary matrices $u$ and $v$, and likewise the eigen-decomposition is taken across a horizontal partition of the gate $e^{-\tau h}$,
\begin{equation} \label{eq:horz}
{e^{ - \tau h}} = \left( {w\sqrt d } \right)\left( {\sqrt d {w^\dag }} \right),
\end{equation}
see Fig.\ref{fig:PathIntegral}(d). Here $w$ is a unitary matrix, which follows from $e^{-\tau h}$ being Hermitian, and $d$ are the eigenvalues [which can be argued to be strictly positive for sufficiently small time step $\tau$]. The SVD and eigen-decompositions are performed throughout the network according to the pattern indicated in Fig.\ref{fig:PathIntegral}(a), and a new square network of tensors $A$, tilted $45^\circ$ with respect to the original, is formed by contracting groups of the resulting tensors together as indicated Fig.\ref{fig:PathIntegral}(b-c). 

In step (ii) the network of tensors $A$ is then coarse-grained in the Euclidean time direction using standard techniques, i.e. by combining pairs of rows together and then truncating the resulting squared bond index similar to the HOTRG\cite{TRG3D} method [see Appendix \ref{app:Compress} for details], until the network is sufficiently isotropic in terms of its correlations. One way to examine how close the network is to being isotropic is to compute the spectra of the transfer matrices formed by tracing out the horizontal or vertical indices of a single $A$ tensor, whose decay should match as closely as possible.

\section{Coarse-graining tensor networks} \label{sect:CG}
Consider a tensor network $\G$ consisting of copies of four index tensors $A_{ijkl}$ that we assume are arranged in an $L\times L$ square-lattice network with periodic boundary conditions. Our goal is to contract this network, or perhaps this network with single or multiple impurity tensors, to evaluate the scalar, denoted $\bk{\G}$, associated to network. As an exact contraction of the network $\G$ is exponentially expensive in system size $L$ one must rely on approximations in order to evaluate a large network. In this section we first describe the generic concept of \emph{local approximations} that could be employed to approximate such a contraction, then discuss the class of local approximation used in TRG, namely the truncated singular value decomposition (SVD), before introducing the particular class of local approximation that the TNR algorithm is based on, which we call \emph{projective truncations}.

\subsection{Local approximations} \label{sect:Local}
\begin{figure}[!t!b]
\begin{center}
\includegraphics[width=8.5cm]{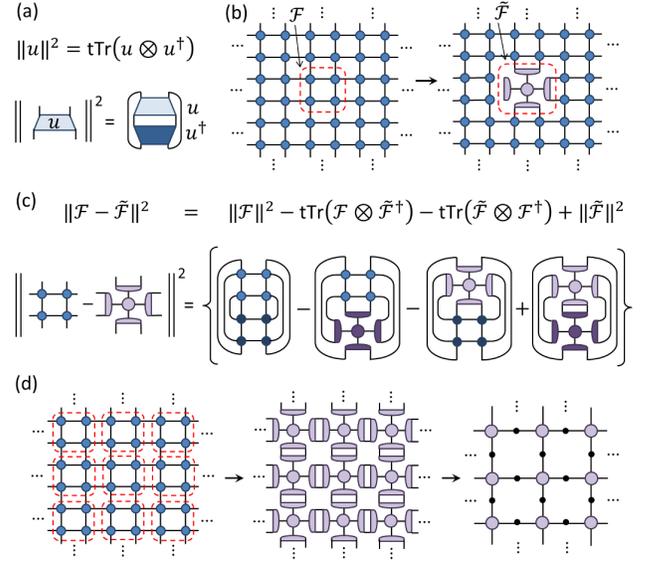}
\caption{(a) Depiction of [the square of] the Hilbert-Schmidt norm of a four index tensor $u$. Note that a darker shade is used to represent the conjugate tensor, which is also drawn with opposite vertical orientation. (b) Given a square lattice tensor network, we wish to replace a $2\times 2$ block of tensors $\F$ from the network with a different sub-network of tensors $\tF$. (c) The square of the difference between $\F$ and $\tF$ under the Hilbert-Schmidt norm is depicted, where again darker shades are used to depict conjugate tensors which are drawn with opposite vertical orientation to regular tensors. The replacement in (b) is valid if the difference $\bnorm{ {\F} - {\tF}}$ is sufficiently small. (c) Assuming the local the square-lattice network is homogeneous, one can replace $\F$ by $\tF$ in all $2\times 2$ blocks. A coarser square-lattice network is obtained after contraction between pairs of three index tensors.}
\label{fig:LocalApprox}
\end{center}
\end{figure}

Let $\F$ denote a sub-network of tensors, for example a $2 \times 2$ block of tensors $A$, from the full network $\G$. The key idea underlying coarse-graining methods for tensor networks is that of the \emph{local approximation}: that one can safely replace a sub-network $\F$ with a different network of tensors $\tF$ if they differ by a small amount  $\varepsilon$,
\begin{equation} \label{eq:epsilon}
\varepsilon \equiv \left\| {\F} - {\tF} \right\|, \\
\end{equation} 
where we assume for convenience that $\F$ has been normalized such that $\left\| \F \right\| = 1$. If this condition is fulfilled, then the scalar $\bk{\G}$ associated to the contraction of network $\G$ will only differ by a small amount $O(\varepsilon)$ under replacement of sub-network $\F$ by new sub-network $\tF$. Note that we use the Hilbert – Schmidt norm,
\begin{equation} \label{eq:norm}
\left\| A \right\| = \sqrt {{\text{tTr}}\left( {A\otimes{A^{\dag} }} \right)}, 
\end{equation}
where `tTr' denotes the tensor trace, or equivalently the contraction of all indices, between two tensors of equal dimensions [or two networks with matching `open' indices], see also Fig.\ref{fig:LocalApprox}(a). In general, renormalization schemes for tensor networks, such as TRG or the focus of this manuscript, TNR, employ a pattern local approximations over all positions on the tensor network, in conjunction with contractions, in order to generate coarser networks of tensors. For example, as illustrated in Fig.\ref{fig:LocalApprox}, local replacements of all  $2 \times 2$ blocks of tensors $\F$ with a new network $\tF$, consisting of a four-index core tensor surrounded by three-index tensors, can result in a coarser, $(L/2)\times (L/2)$ network of new four index tensors. Assuming that sufficient accuracy could be maintained over repeated steps, this procedure could be iterated $O(\log_2 L)$ times, resulting in a network of $O(1)$ linear dimension which could then be exactly contracted. 

\subsection{Truncated singular value decomposition} \label{sect:TRG}
In principle, any form of local approximation capable of yielding a small error $\varepsilon$ in Eq.\ref{eq:epsilon} could be viable as part of a coarse-graining scheme. In the original TRG algorithm proposed by Levin and Nave\cite{TRG}, and in many of the generalizations and improvements to TRG\cite{SRG, TRGenv, TRG3D}, the local approximations that are used are based upon a truncated singular value decomposition (SVD) of single tensors [or a generalized form of the SVD known as the higher-order singular value decomposition\cite{HOSVD} (HOSVD)], which we now discuss.

Consider a four index tensor $A_{ijkl}$ where each index is of dimension $\chi$. If the tensor is viewed as a $\chi^2 \times \chi^2$ matrix according to the pairing of indices $A_{[ij][kl]}$ then the SVD can be performed, 
\begin{equation} \label{eq:SVD}
A_{ijkl} = \sum\limits_{m}^{\chi^2} u_{ijm} s_{mm} v_{mkl}, 
\end{equation}
where $u$ and $v$ are unitary according to grouping of indices $u_{[ij][m]}$ and $v_{[m][kl]}$ respectively, and $s$ is a positive diagonal matrix of singular values $\lambda$, i.e. $s_{nm} = \delta_{nm} \lambda_m$, that we assume are ordered such that $\lambda_m < \lambda_{m+1}$. If we truncate the SVD to retain only the $\chi'<\chi^2$ largest singular values then the decomposition becomes approximate,
\begin{equation} \label{eq:SVDtrunc}
A_{ijkl} \approx \sum\limits_{m}^{\chi'} u_{ijm} s_{mm} v_{mkl},
\end{equation}
where the truncation error $\varepsilon$, as defined in Eq.\ref{eq:epsilon}, is seen to equal to the square-root of the sum of the squares of the discarded singular values,
\begin{equation}
\varepsilon  = \sqrt {\sum\limits_{m = \chi'  + 1}^{\chi^2} {\left( {\lambda_m} \right)^2} }.
\end{equation}
Here we have assumed that the tensor $A$ was normalized, $\left\| A \right\| = 1$, or equivalently that the singular values were normalized as $\sqrt{ {\sum\nolimits_m (\lambda_m})^2} = 1$. The SVD is known to provide the optimally accurate decomposition of a tensor $A$ into the product of a pair of tensors [connected by an index of some rank $\chi'$], thus it has proved vitally useful as the foundation for many previous schemes for the renormalization of tensor networks. 

\subsection{Projective truncations} \label{sect:Project}
\begin{figure}[!t!b]
\begin{center}
\includegraphics[width=8.5cm]{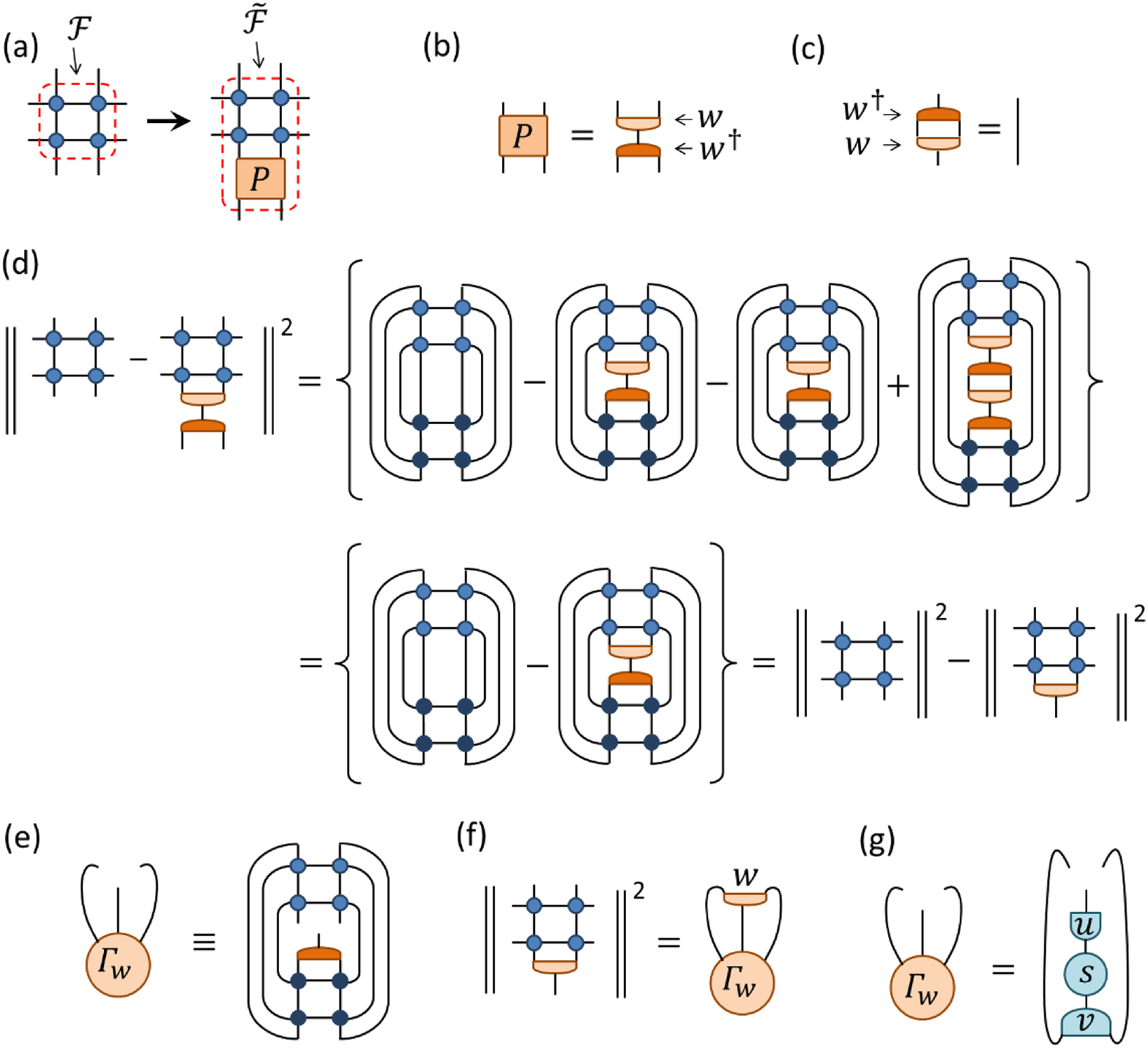}
\caption{(a) In a \emph{projective truncation} a sub-network $\F$ is replaced by a new sub-network $\tF$, which consists of a projector $P$ applied to the original sub-network, i.e. $\tF = \F P$. (b) Here we assume that $P$ is decomposed as a product of an isometric tensor $w$ and its conjugate, $P = w w^{\dag}$. (c) By definition, isometry $w$ contracts to identity with its conjugate, $w^{\dag} w = \mathbb{I}$. (d) The square of the error in a projective truncation is expanded as a sum of four terms; however given that $P^2=P$, two of the terms cancel, see also Eq.\ref{eq:proj3}. (e)  The environment $\Gamma_w$ of isometry $w$ is defined as the network that results by removing a single instance of $w$ from ${\left\| \F w \right\|}^2$, see also Eq.\ref{eq:proj5}. (f) By construction, the contraction of $w$ and its environment $\Gamma_w$ is equal to ${\left\| \F w \right\|}^2$. (g) Environment $\Gamma_w$ is decomposed, via singular value decomposition (SVD), into a product of isometric tensors $u$, $v$ and diagonal matrix $s$.}
\label{fig:ProjApprox}
\end{center}
\end{figure}

\begin{figure}[!t!b]
\begin{center}
\includegraphics[width=8.5cm]{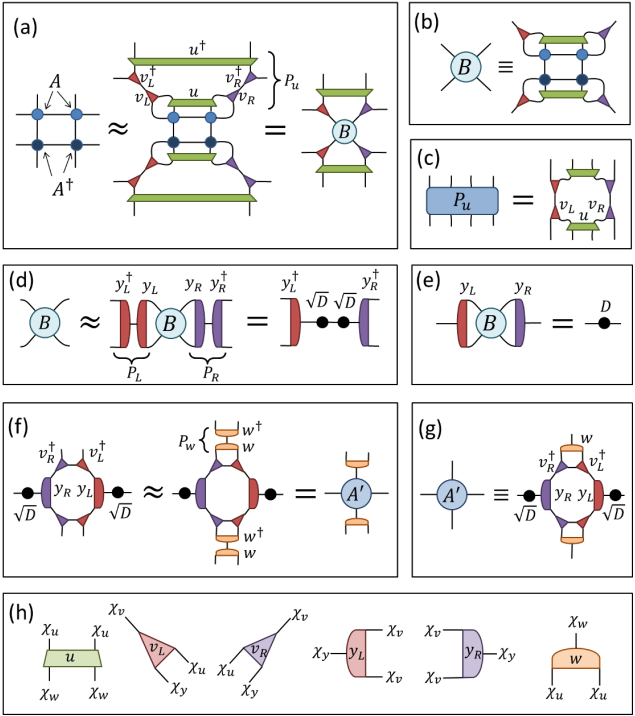}
\caption{(a) Details of the projective truncation made at the first step of the TNR iteration; here two copies of a projector $P_u$, which is composed of a product isometric and unitary tensors, are applied to a $2\times 2$ block of $A$ tensors. (b) Definition of four index tensor $B$. (c) Projector $P_u$ is formed from isometries $v_L$, $v_R$  and disentangler $u$ [and their conjugates]. (d) Details of the projective truncation made at the second step of the TNR iteration. (e) Definition of matrix $D$. (f) Details of the projective truncation made at the third step of the TNR iteration. (g) Definition of new four-index tensor $A'$, copies of which comprise the coarse-grained square-lattice tensor network. (h) Delineation of the different dimensions $\{\chi_u, \chi_v, \chi_w, \chi_y \}$ of indices on tensors $\{u, v_L, v_R, y_L, y_R, w\}$.}
\label{fig:TNRsteps}
\end{center}
\end{figure}

The TNR approach requires use of a broader class of local approximation than those based upon the SVD, which we term \emph{projective truncations}. In a projective truncation, a local sub-network $\F$ is replaced by a new local network $\tF$ that consists of a projector $P$, which satisfies $PP^{\dag} = P^2  = P$, acting on some or all of the open indices of $\F$,
\begin{equation} \label{eq:proj1}
\tF = \F P  = \F w w^{\dag}, 
\end{equation}
see Fig.\ref{fig:ProjApprox}(a) for an example. Here we have expanded the projector $P$ as the product of an isometric tensor $w$ and its conjugate, $P = w w^{\dag}$, where the isometry satisfies $w^{\dag} w = \mathbb{I}$, see also Fig.\ref{fig:ProjApprox}(b-c). [Note that, as part of the TNR algorithm, we shall also consider cases where projector is decomposed as a more complicated product of many different isometries, see for example Fig.\ref{fig:TNRsteps}(c)]. Projective truncations are a particularly useful class of local approximation, as the figure of merits to optimize projector $P$ takes a very simple form. To see this, we first expand the terms in Eq.\ref{eq:epsilon},
\begin{equation} \label{eq:proj2}
\varepsilon^2 = \bnorm{\F} + \bnorm{\tF} - {\text{tTr}}\left( {\F \otimes {{\tilde \F}^*}} \right) - {\text{tTr}}\left( {\tilde \F \otimes {\F^*}} \right) \hfill .
\end{equation}
Notice that, as $PP^{\dag}  = P$, for a projective truncation we have, 
\begin{equation} \label{eq:proj3}
\bnorm{\tF}  = {\text{tTr}}\left( {\F \otimes {{\tilde \F}^*}} \right) = {\text{tTr}}\left( {\tilde \F \otimes {\F^*}} \right).
\end{equation}
It follows that the expression for the replacement error $\varepsilon$ can be simplified, 
\begin{align} \label{eq:proj4}
\varepsilon &= \sqrt{{\left\| \F \right\|^2}  - {\left\| \F P \right\|^2} }\nonumber \\ 
&= \sqrt{{\left\| \F \right\|^2}  - {\left\| \F w \right\|^2} },
\end{align}
where we have again made use of $PP^{\dag} = P$ in reaching the second line of working, see also Fig.\ref{fig:ProjApprox}(d). 

Let us now turn to the problem of optimizing the projector $P$ such that the error $\varepsilon$ of Eq.\ref{eq:proj4} is minimized. For simplicity we consider the case where the projector $P$ decomposes as a product of a single isometry $w$ and its conjugate, $P = w w^{\dag}$, although a key feature of the method we discuss is that it can be applied to the more complicated case, an in Fig.\ref{fig:TNRsteps}(c), where $P$ is represented as a product of several different isometric tensors. Notice that the error $\varepsilon$ of Eq.\ref{eq:proj4} is minimized when ${\left\| \F w \right\|} $ is maximized, which follows as ${\left\| \F w \right\|} \le {\left\| \F \right\|}$; we now discuss an iterative strategy for optimizing isometry $w$ to maximize the expression ${\left\| \F w \right\|} $. 

The strategy we employ is based upon \emph{linearization} of the cost function. Given that the expression ${\left\| \F w \right\|}^2$ has quadratic dependence on isometry $w$ [or, more specifically, in depends on both $w$ and $w^\dag$], we simplify the problem by temporarily holding $w^\dag$ fixed and then solving the resulting linear optimization for $w$, and iterating these steps until $w$ is sufficiently converged. Note that this follows the same strategy employed to optimize tensors in a MERA as described in Ref.\onlinecite{Algorithms}, to which we refer the interested reader for more details. We begin by expressing the closed network ${\left\| \F w \right\|}^2$ in a factorized form,
\begin{equation} \label{eq:proj5}
{\left\| \F w \right\|}^2 = \textrm{tTr} \left( \Gamma_w \otimes w \right), 
\end{equation}
where $\Gamma_w$, referred to as the environment of $w$, represents the contraction of everything in ${\left\| \F w \right\|}^2$ excluding tensor $w$, see also Fig.\ref{fig:ProjApprox}(e-f). The singular value decomposition of the environment ${\Gamma _w}$, when considered as a matrix according to the same partition of indices for which tensor $w$ is isomeric, is then taken, 
\begin{equation} \label{eq:proj6}
{\Gamma _w} = us{v^\dag }, 
\end{equation}
as shown in Fig.\ref{fig:ProjApprox}(g). The isometry $w$ is then updated to become,
\begin{equation} \label{eq:proj7}
w = v{u^\dag },
\end{equation}
where it can be argued that this choice of updated isometry maximizes Eq.\ref{eq:proj5}. However, as the cost function was linearized, such that $w^\dag$ in the environment $\Gamma_w$ was held fixed, these steps of (i) computing the environment $\Gamma _w$ and (ii) obtaining an updated $w$ through the SVD of the environment $\Gamma _w$ must be iterated many times, until the solution converges. 

\section{Tensor network renormalization} \label{sect:TNR}
Tensor network renormalization is a class of coarse-graining scheme designed to be compatible with proper removal of all short-ranged correlations at each RG step\cite{TNR}. For any given lattice geometry there are many potential TNR schemes the fulfill this requirement. In this manuscript we focus on a particular implementation of TNR for a $2D$ square lattice network that we call the \emph{binary} TNR scheme, as introduced in Ref.\onlinecite{TNR}, which reduces the linear dimension of the network by a factor of $2$ with each RG step. In Appendix \ref{app:Ternary} we discuss a \emph{ternary} TNR scheme, which reduces the linear dimension of the network by a factor of $3$ with each RG step, and in Appendix \ref{app:Isotropic} we discuss an \emph{isotropic} binary TNR scheme that treats both dimensions of the tensor network equally while also reducing the linear dimension of the network by a factor of $2$ at each RG step. Similarly TNR can also be implemented in other lattice geometries besides the square-lattice, including those in higher dimensions. The majority of the algorithmic details we present for implementation of the binary TNR scheme carry over to other TNR schemes.

\begin{figure}[!t!b]
\begin{center}
\includegraphics[width=8.5cm]{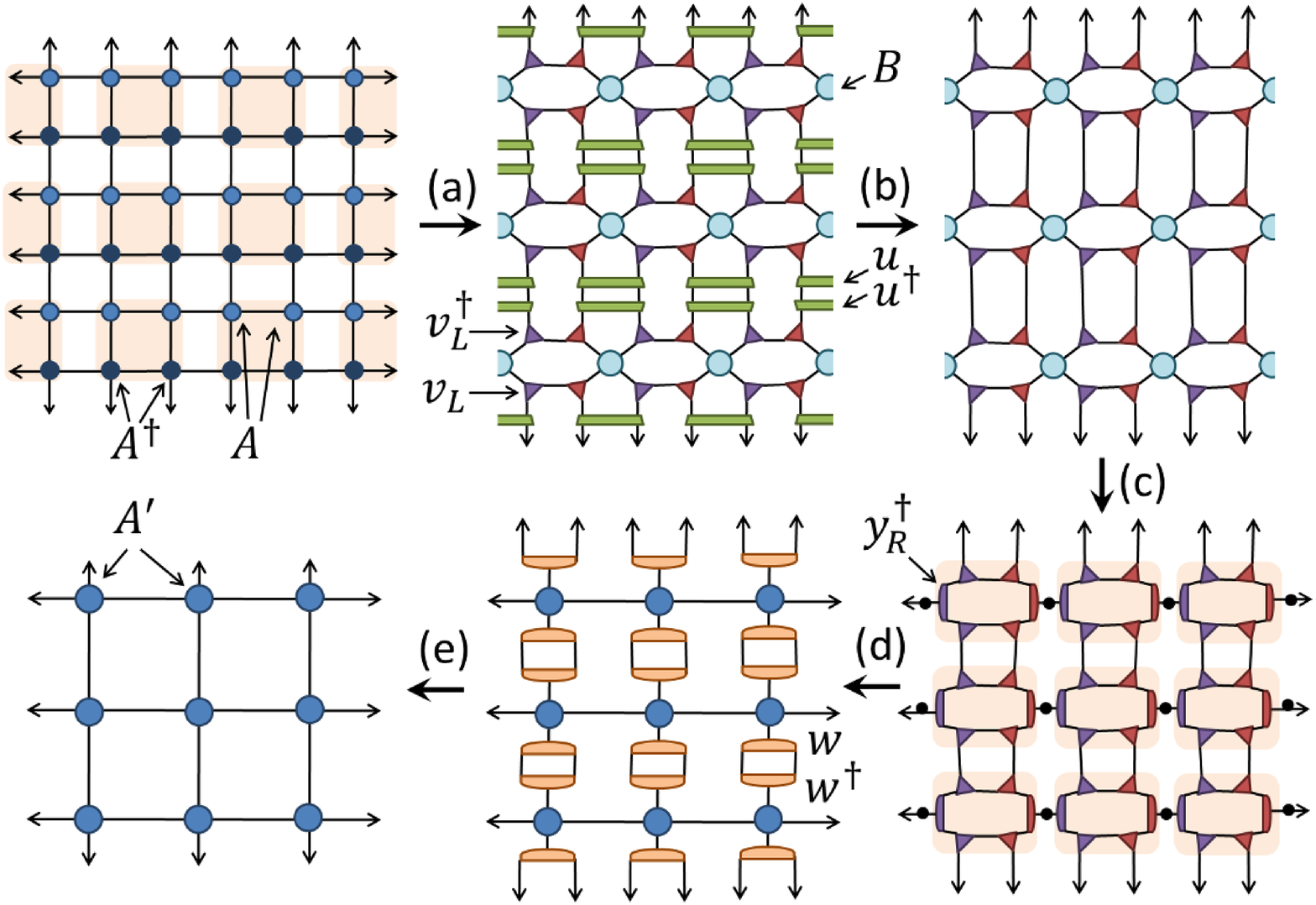}
\caption{The sequence of coarse-graining steps used in the binary TNR scheme in order to map an initial square lattice of tensors $A$, where every second row of tensors has been conjugated as described in Appendix \ref{app:Reflect}, to a coarser square lattice composed of tensors $A'$. (a) A projective truncation is made on all $2\times 2$ blocks of tensors, see Fig.\ref{fig:TNRsteps}(a-c). (b) Conjugate pairs of disentanglers $u$ are contracted to identity. (c) A projective truncation is made on all $B$ tensors, see Fig.\ref{fig:TNRsteps}(d-e). (d) A final projective truncation is made, see Fig.\ref{fig:TNRsteps}(f-g) for details. (e) Conjugate pairs of isometries $w$ are contracted to identity.}
\label{fig:CoarseGrain}
\end{center}
\end{figure}
 
\subsection{Coarse-graining step of the binary TNR scheme} \label{sect:binTNR}
The starting point for an iteration of the binary TNR scheme is a square lattice tensor network $\G$ composed of four index tensors $A_{ijkl}$, where indices are assumed to be of some dimension $\chi$. As discussed in Sect.\ref{sect:TN} such a network could represent the partition function of a $2D$ classical statistical system or the Euclidean path integral of a $1D$ quantum system. For simplicity we assume that network $\G$ is spatially homogeneous, i.e. that all its tensors are copies of a unique tensor $A_{ijkl}$, while noting that the algorithm can easily be extended to deal with non-homogeneous networks [special examples, including networks with an open boundary or a defect line, can be handled using similar methods as those developed in the context of MERA\cite{BoundMERA1, BoundMERA2, MinUpdate, ImpurityMERA1, ImpurityMERA2}, and are discussed separately in Ref.\onlinecite{TNRimpurity}]. We also assume that $\G$ is invariant under complex conjugation plus reflection on the horizontal axis, as discussed further in Appendix \ref{app:Reflect}. Again, this assumption is not strictly necessary, but is useful in simplifying the TNR algorithm. In the case that $\G$ represents a Euclidean path integral of a quantum Hamiltonian $H$ the presence of this symmetry follows from $H$ being Hermitian, while in the case that $\G$ represents the partition function of a classical system the symmetry is present if the underlying $2D$ classical statistical model has an axis with which it is invariant under spatial reflection. We now describe the coarse-graining steps involved in an iteration of the binary TNR scheme, which maps network $\G$ to the coarser network $\G'$ whose linear dimension has been reduced by a factor of 2, before discussing the optimization of the tensors involved and other algorithmic components in more detail. 

The first step of the iteration is to apply a particular gauge change on the horizontal indices on every second row of tensors in $\G$, as discussed in Appendix \ref{app:Reflect}. Here the gauge change is chosen such that it is equivalent to flipping top-bottom indices of tensors $A$ and taking the complex conjugation, as such we denote the transformed tensors $A^\dag$. That such a gauge transformation exists follows from the assumed reflection symmetry. Next, Fig.\ref{fig:CoarseGrain} depicts the remaining steps in transforming network $\G$ into the coarser network $\G'$. In Fig.\ref{fig:CoarseGrain}(a), a projective truncation is enacted on $2\times 2$ blocks of tensors $A$ [where two of the tensors are have undergone the aforementioned change of gauge as $A^\dag$], the details of which are shown in Fig.\ref{fig:TNRsteps}(a). The projector $P_u$ used at this step is represented as a product of two isometries $v_L$ and $v_R$ and a unitary tensor $u$ [and their conjugates] as shown in Fig.\ref{fig:TNRsteps}(c). The unitary tensors $u$, which we call \emph{disentanglers}, act on two neighboring indices such that, if we regard each index of the network as hosting a $\chi$-dimensional complex vector space $\mathbb{V}_\chi$, they describe a mapping between vector spaces,   
\begin{equation} \label{eq:u}
u: \mathbb{V}_\chi \otimes \mathbb{V}_\chi \rightarrow \mathbb{V}_\chi \otimes \mathbb{V}_\chi.
\end{equation}
By virtue of being unitary, the disentanglers satisfy $u^{\dagger} u=\mathbb{I}^{\otimes 2}$ where $\mathbb{I}$ is the identity operator on $\mathbb{V}_\chi$. Conceptually, the role of disentanglers is to remove short-range correlations that would otherwise be missed, as discussed in greater detail in Ref.\onlinecite{TNR}, and they constitute the key difference between TNR and previous tensor renormalization schemes. Isometries $v_L$ and $v_R$ each map two indices in the network, one horizontal and one vertical, to a new index of some chosen dimension $\chi' \le \chi^2$,
\begin{equation} \label{eq:v}
v_L: \mathbb{V}_{\chi'} \rightarrow \mathbb{V}_{\chi} \otimes \mathbb{V}_\chi,\; \; \; \; v_R: \mathbb{V}_{\chi'} \rightarrow \mathbb{V}_{\chi} \otimes \mathbb{V}_\chi,
\end{equation}
where the new index has been regarded as hosting a $\chi'$-dimensional complex vector space $\mathbb{V}_{\chi'}$. By definition, isometries satisfy $v_L^{\dagger} v_L = v_R^{\dagger} v_R = \mathbb{I'}$, with $\mathbb{I}'$ the identity operator on $\mathbb{V}_{\chi'}$. After the coarse-graining step of Fig.\ref{fig:CoarseGrain}(a), it is useful to define a new four index tensor $B$, which is defined from the block of $A$ tensors and from $u$, $v_L$ and $v_R$, as depicted in Fig.\ref{fig:TNRsteps}(b).

After the projective truncation implemented by projector $P_u$ is enacted on all $2\times 2$ blocks of tensors, pairs of disentanglers from neighboring blocks can annihilate to identity, see Fig.\ref{fig:CoarseGrain}(b), leaving a network of $B$ tensors interspersed with groups of isometries $v_L$ and $v_R$. Next, as depicted in Fig.\ref{fig:CoarseGrain}(b) and further detailed in Fig.\ref{fig:TNRsteps}(d-e), a projective truncation is made on $B$ tensors. Two projectors $P_L$ and $P_R$ are used at this step, acting on the left or right indices of $B$ tensor respectively, each formed as a product of isometries, $P_L \equiv y_L y_L^{\dagger}$ and $P_R \equiv y_R y_R^{\dagger}$. Isometries $y_L$ and $y_R$, which satisfy $y_L^{\dagger} y_L = y_R^{\dagger} y_R = \mathbb{I}'$, each map two indices to a single index also assumed to be of dimension $\chi'$,
\begin{equation} \label{eq:y}
y_L: \mathbb{V}_{\chi'} \rightarrow \mathbb{V}_{\chi'} \otimes \mathbb{V}_{\chi'},\; \; \; \; y_R: \mathbb{V}_{\chi'} \rightarrow \mathbb{V}_{\chi'} \otimes \mathbb{V}_{\chi'}.
\end{equation}
We then define matrix $D$ from enacting isometries $y_L$ and $y_R$ on tensor $B$, as shown Fig.\ref{fig:TNRsteps}(e). It is useful, though not strictly necessarily, to work in a gauge where matrix $D$ is diagonal and positive, which can always be achieved through proper choice of gauge on isometries $y_L$ and $y_R$, such that the matrix can easily be decomposed as $D = \sqrt{D} \sqrt{D}$.

After projective truncations have been made on $B$ tensors a final projective truncation, as shown Fig.\ref{fig:CoarseGrain}(d) and further detailed in Fig.\ref{fig:TNRsteps}(f-g), is made using projector $P_w \equiv w w^{\dagger} $ with isometry $w$ mapping two $\chi$-dimensional indices to a single index of dimension $\chi'$, 
\begin{equation} \label{eq:w}
w: \mathbb{V}_{\chi'} \rightarrow \mathbb{V}_\chi \otimes \mathbb{V}_\chi,
\end{equation}
where the isometry satisfies $w^{\dagger} w= \mathbb{I'}$. After projector $P_w$ has been implemented throughout the network, pairs of isometries $w$ from neighboring cells can annihilate to identity with their conjugates, as depicted in Fig.\ref{fig:CoarseGrain}(e). This final step yields a coarse-grained square lattice $\G'$ of four index tensors $A'$, whose indices are of some specified dimension $\chi' \le \chi^2$. Notice that the new four-index tensor $A'$ is defined from a product of various tensors obtained throughout the coarse-graining step, namely from $\{v_L, v_R, y_L, y_R, \sqrt{D}, w\}$, as shown Fig.\ref{fig:TNRsteps}(g). 

\begin{figure}[!t!b]
\begin{center}
\includegraphics[width=8cm]{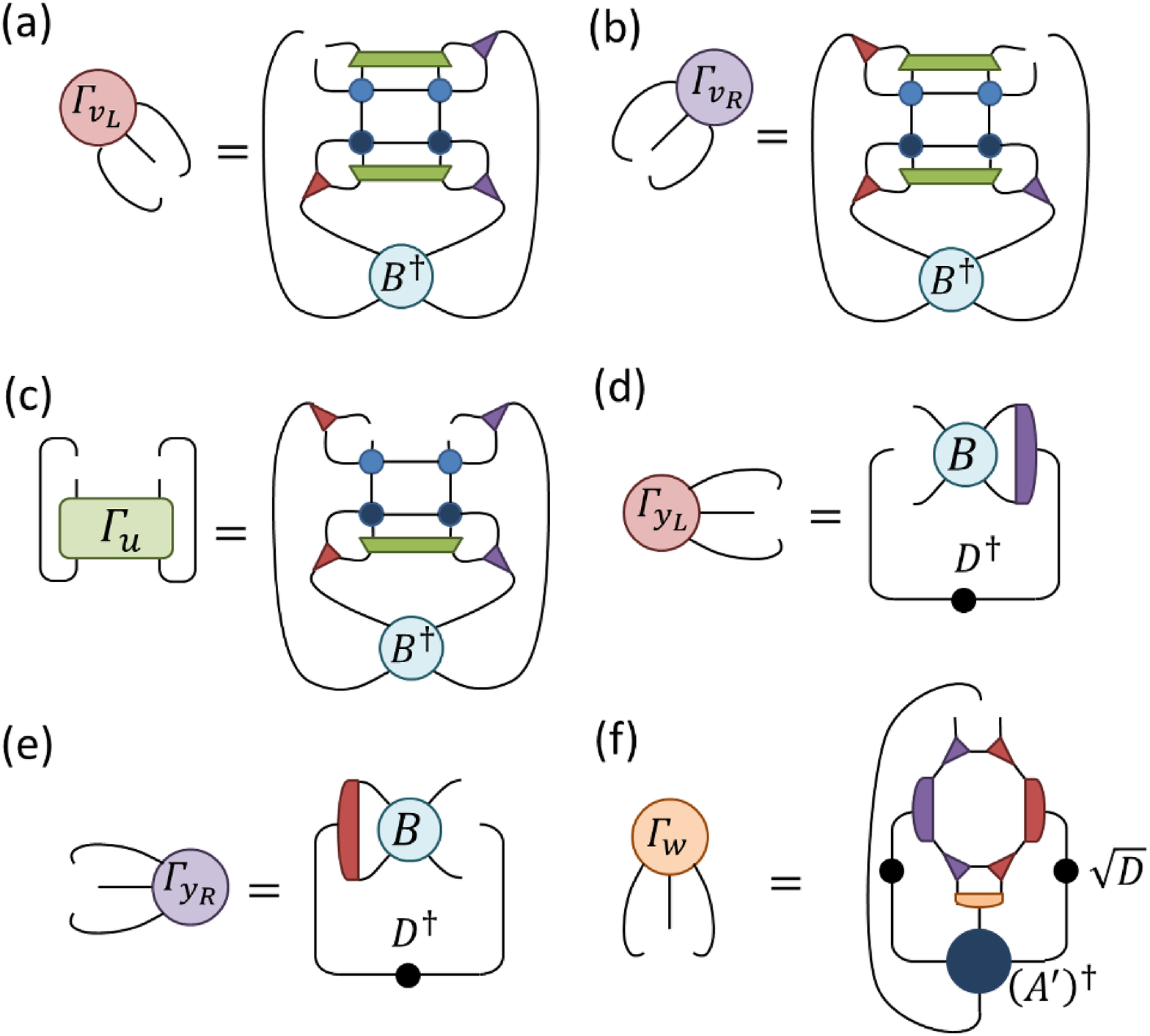}
\caption{The linearized environments of tensors $\{v_L, v_R, u, y_L, y_R, w\}$ involved in an iteration of the binary TNR scheme. (a-c) Environments $\Gamma_{v_L}$, $\Gamma_{v_R}$ and $\Gamma_{u}$ of the isometries $v_L$, $v_R$ and disentangler $u$ involved in the first projective truncation of the TNR iteration, as detailed in Fig.\ref{fig:TNRsteps}(a). (d-e) Environments $\Gamma_{y_L}$ and $\Gamma_{y_R}$ of isometries $y_L$ and $y_R$ from the second projective truncation of the TNR iteration, as detailed in Fig.\ref{fig:TNRsteps}(d). (f) Environment $\Gamma_{w}$ of isometry $w$ from the third projective truncation of the TNR iteration, as detailed in Fig.\ref{fig:TNRsteps}(f).}
\label{fig:Optimization}
\end{center}
\end{figure}

\subsection{Optimization of tensors} \label{sect:Opt}
Each coarse graining iteration of the binary TNR scheme follows from a series of projective truncations. The projectors involved can be optimized using the iterative SVD update strategy, described in Sect.\ref{sect:Project}, as we now discuss in more detail.

The first step of the TNR iteration, as depicted Fig.\ref{fig:CoarseGrain}(a), requires optimization of a projector $P_u$ composed of isometries $v_L$ and $v_R$ and disentangler $u$, as shown Fig.\ref{fig:TNRsteps}(c). To update one of these tensors one first computes its environment, where the environments $\Gamma_{v_L}$, $\Gamma_{v_R}$ and $\Gamma_{u}$ are shown in Fig.\ref{fig:Optimization}(a-c), then updates the tensor from the SVD of the environment as discussed in Sect.\ref{sect:Project}. Each of the tensors $v_L$, $v_R$, $u$ should be updated in turn and the process iterated until all tensors are sufficiently converged [which typically requires of order a few hundred iterations]. The computational cost of computing each of the environments scales as $O(\chi^7)$, assuming the indices involved in the network are all $\chi$-dimensional.

The second projective truncation step of the TNR iteration, as depicted Fig.\ref{fig:CoarseGrain}(c), requires the optimization of isometries $y_L$ and $y_R$, which again are optimized through alternating, iterative SVD updates based upon calculation of their environments $\Gamma_{y_L}$ and $\Gamma_{y_L}$, as shown in Fig.\ref{fig:Optimization}(d,e). Note that, in order to ensure that the reflection symmetry [which was assumed to be present in the initial tensor network] is preserved, it is necessary to symmetrize the environments before performing the SVD, as described in Appendix \ref{app:Reflect}. The computational cost of computing each of the environments is $O(\chi^5)$ assuming the indices involved in the network are $\chi$-dimensional. 

The third and final projective truncation of the TNR iteration, as depicted Fig.\ref{fig:CoarseGrain}(d), requires optimization of isometry $w$ which can be achieved through iterative SVD updates of the environment $\Gamma_w$, depicted in Fig.\ref{fig:Optimization}(f). The computational cost of computing the environment $\Gamma_w$ is $O(\chi^6)$ assuming the indices involved in the network are $\chi$-dimensional.

\subsection{RG flow of tensors} \label{sect:RG}
The single iteration of the TNR approach described in Sect.\ref{sect:binTNR}, which mapped the initial tensor network $\G$ to the coarser network $\G '$, can be iterated many times to generate a sequence of increasing coarse-grained networks,
\begin{equation} \label{eq:TNflow}
{\G^{(0)}} \to {\G^{(1)}} \to {\G^{(2)}} \to  \cdots  \to {\G^{(s)}} \to  \cdots 
\end{equation}
with $\G^{(0)} \equiv \G$ now as the initial network and network $\G^{(s)}$ as tensor network after $s$ iterations of TNR, where the linear dimension of the lattice $\G^{(s)}$ has been reduced by a factor of 2 compared to the previous network $\G^{(s-1)}$. Each network $\G^{(s)}$ consists of copies of a four index tensor $A_{ijkl}^{(s)}$, whose indices are of dimensions $\chi^{(s)}$, arranged in a square lattice configuration. Here the initial dimension $\chi^{(0)}$ is fixed from the starting tensors $A_{ijkl}^{(0)}$, while the dimensions $\chi^{(s)}$ at later steps are user specified, and can be increased to improve the accuracy of the calculation at the cost of increasing the computational expense [aspects of computational efficiency will further be discussed in Sect.\ref{sect:Alg}]. Tensors $A^{(s+1)}$ are defined from [four copies of] the previous tensor $A^{(s)}$ under coarse-graining via a product of optimized isometric tensors, 
\begin{equation} \label{eq:Asingle}
{A^{(s)}} \xrightarrow{\; \left\{v_L^{(s)},v_R^{(s)},{u^{(s)}},y_L^{(s)},y_R^{(s)},{w^{(s)}} \right\} \;}{A^{(s + 1)}},
\end{equation}
as depicted in Fig.\ref{fig:TNRsteps}(g), such that we can consider TNR as generating an RG flow in the space of [four-index] tensors,
\begin{equation} \label{eq:Aflow}
{A^{(0)}} \to {A^{(1)}} \to {A^{(2)}} \to  \cdots  \to {A^{(s)}} \to  \cdots. 
\end{equation}
Properties of the system under consideration, such as the expectation values of local observables, can be computed from both the coarse-grained tensors ${A^{(s)}}$ and the tensors $\{u^{(s)},v_L^{(s)}, v_R^{(s)}, y_L^{(s)}, y_R^{(s)}, w^{(s)}\}$ involved in the coarse-graining, as further discussed in Sect.\ref{sect:Observables}.
 
\subsection{Algorithmic details} \label{sect:Alg}
While the TNR algorithm discussed thus far is a viable implementation of approach, we now detail how this basic TNR algorithm can be improved in a number of ways.    

In most calculations it is convenient to use the same bond dimensions $\chi^{(s)}$ for all RG steps $s$, i.e. such that $\chi^{(s)} = \chi$ for $s > 0$ [where $\chi^{(0)}$ is fixed by the local dimension of the model under consideration]. However, even when using the same bond dimension throughout different coarse-graining iterations, it can still be useful, in order to maximize the efficiency of the TNR approach, to use different bond dimensions on vertical and horizontal indices of tensors $A^{(s)}$, as well as different dimensions on tensors that appear at intermediate steps of each coarse-graining iteration. Following this idea, four different refinement parameters $\{\chi_u, \chi_v, \chi_w, \chi_y \}$ can be defined, each denoting a dimension of an outgoing index on different isometric tensors as indicated in Fig.\ref{fig:TNRsteps}(h). Here $\chi_w$ and $\chi_y$ denote the dimensions of vertical and horizontal indices on $A$ tensors respectively, while $\chi_u$ and $\chi_v$ denote dimensions of indices that only appear at intermediate steps of each TNR iteration. The ratio of dimensions should be adjusted such that the truncation errors $\varepsilon$ given at different intermediate steps become roughly equal; in practice such dimensions can be determined heuristically. For several test models it has been observed that the different optimal bond dimensions are, to good approximation, linearly related to one another [i.e. the ratios of different optimal bond dimensions remain roughly constant for a given model]. This implies that the cost scaling of an algorithmic step can be specified unambiguously, for instance as having cost $O(\chi^7)$, without detailing the dependence on the different dimensions involved [as they are linearly related]. Note that, in the benchmark results presented in Sect.\ref{sect:Bench} we label results of a TNR simulation by a single dimension $\chi$, by which we mean the \emph{largest} of the bond dimensions $\{\chi_u, \chi_v, \chi_w, \chi_y \}$ that was used in the simulation.

The computational cost of the binary TNR can be reduced by implementing an additional projective truncation on the $2\times 2$ blocks of $A$ tensors at the start of the TNR iteration, as described in Appendix \ref{app:Reduce}. This additional step reduces the cost of computing the environments $\Gamma_{v_L}$, $\Gamma_{v_R}$ and $\Gamma_{u}$, see Fig.\ref{fig:Optimization}(a-c), from $O(\chi^7)$ to $O(\chi^6)$, thus also reduces the cost of optimizing tensors $v_L$, $v_R$ and $u$. When utilizing the ideas of Appendix \ref{app:Reduce} the tensor contractions required to implement TNR are all of cost $O(\chi^6)$ or less, thus the efficiency of the algorithm is greatly improved. 

The accuracy of TNR, for a fixed bond dimension $\chi$, can be improved by taking a larger environment into account for optimization of the tensors, using an approach similar to the approach of Refs.\onlinecite{TRGenv,SRG} for improving standard TRG by taking the environment into consideration. Appendix \ref{app:Env} describes how the tensors involved in the TNR coarse-graining iteration can be optimized using a larger [though still local] environment, and the benefits of doing so. Note that, by also incorporating the ideas of Appendix \ref{app:Reduce}, the leading order computational cost of optimizing using the larger environments remains unchanged at $O(\chi^6)$. It is also possible to modify TNR to take account of the full environment from the network, similar to how SRG\cite{SRG} modifies the TRG to take account of the full environment. This modification, which we shall not detail in the present manuscript, requires sweeping the optimization back and forth over different scales of coarse-graining, functioning similarly to the energy minimization algorithm for optimizing a MERA\cite{Algorithms}.

\section{Calculation of observables} \label{sect:Observables}
In this section we discuss how TNR can be applied to compute expectation values of local observables in classical statistical or quantum many-body systems. There are many different ways of performing this calculation; one approach could be to construct a MERA from the TNR coarse-graining transformations, as described in Ref.\onlinecite{TNRtoMERA}, from which expectation values could then be computed using standard MERA techniques\cite{Algorithms}. In this manuscript we describe a different approach based upon directly coarse-graining the network with the addition of an impurity tensor. 

\begin{figure}[!t!b]
\begin{center}
\includegraphics[width=8.5cm]{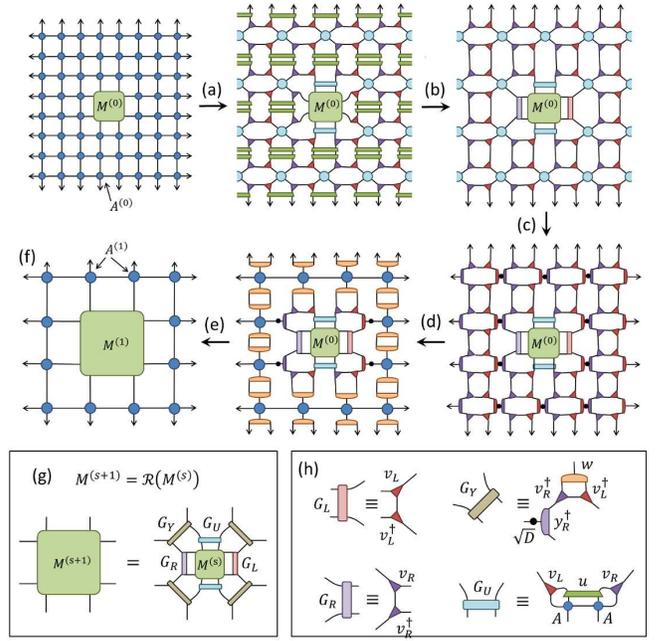}
\caption{(a-e) An iteration of the TNR coarse-graining transformation for a square-lattice tensor network that is homogeneous everywhere except for a $2\times 2$ region that has been replaced with an impurity tensor $M^{(0)}$. The projective truncations of the TNR iteration are applied as usual, see Fig.\ref{fig:CoarseGrain}, neglecting those which are incompatible impurity tensor. (f) The coarse-grained square lattice is homogeneous everywhere except for a $2\times 2$ region occupied by new impurity tensor $M^{(1)}$. (g) The coarse-grained impurity tensor $M^{(s+1)}$ is given by enacting ascending superoperator $\R$, defined from two body gates $\{ G_R, G_L, G_U, G_Y \}$, on the impurity tensor $M^{(s)}$. (h) Definition of two body gates $\{ G_R, G_L, G_U, G_Y \}$.}
\label{fig:LogSingle}
\end{center}
\end{figure}

\begin{figure}[!t!b]
\begin{center}
\includegraphics[width=8cm]{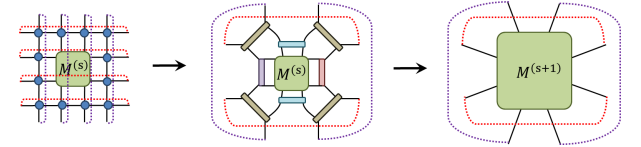}
\caption{A finite network with periodic boundaries and a local impurity $M^{(s)}$ is coarse-grained into a single impurity tensor $M^{(s+1)}$ under the transformation depicted in Fig.\ref{fig:LogSingle}. The expectation value of the local observable corresponding to the impurity is given from the trace of $M^{(s+1)}$ as depicted.}
\label{fig:LogFinite}
\end{center}
\end{figure}

Let $\G^{(0)}$ be a homogeneous square lattice tensor network with periodic boundaries that consists of an $L\times L$ array of copies of a four index tensor $A^{(0)}$. Assume that the sequence of $T = \log_2(L/2)$ TNR transformations have been optimized as to generate a sequence of coarser lattices, see Eq.\ref{eq:TNflow}, where $\G^{(T)}$ is a $2\times 2$ lattice of tensors $A^{(T)}$ that can be exactly contacted. In order to evaluate the expectation value of a local observable, we must now evaluate the tensor network $\G^{(0)}$ with the addition of an impurity tensor representing the local observable under consideration. Here we consider an impurity tensor $M^{(0)}$ that replaces a $2\times 2$ block of $A$ tensors from $\G^{(0)}$. To evaluate the impurity network we use the same projective truncations as was used to coarse-grain the homogeneous network $\G^{(0)}$ everywhere except in the immediate vicinity of the impurity, the presence of which is incompatible with the coarse-graining steps used for the homogeneous system, see Fig.\ref{fig:LogSingle}(a-e). After an iteration of TNR to the impurity network, one obtains a new impurity network, equal to the homogeneous network $\G^{(1)}$ except where a new impurity $M^{(1)}$ replaces a $2\times 2$ block of tensors $A^{(1)}$, see Fig.\ref{fig:LogSingle}(f). The coarse-grained impurity $M^{(1)}$ is defined from applying a set of two-body gates, $\{G_R, G_L, G_U, G_Y\}$, to the initial impurity tensor $M^{(0)}$, as depicted in Fig.\ref{fig:LogSingle}(g), where the gates are functions of $A^{(0)}$ and the isometries $\{ v_L, v_R, u, y_L, y_R, w \}$ used in the TNR iteration, as depicted in Fig.\ref{fig:LogSingle}(h). The coarse-graining transformation can be applied to the impurity network multiple times as to generate a sequence of impurity tensors,
\begin{equation} \label{eq:ImpFlow}
{M^{(0)}} \to {M^{(1)}} \to  \cdots  \to {M^{(T)}},
\end{equation}
each imbedded in an increasingly coarse-grained square-lattice network. We call the transfer operator that maps impurities tensors from one length-scale to the next, which was introduced in Ref.\onlinecite{TNRCFT}, the \emph{ascending superoperator} $\R$, 
\begin{equation} \label{eq:Ascend}
{M^{(s + 1)}} = \R \left( {{M^{(s)}}} \right),
\end{equation}
as shown in Fig.\ref{fig:LogSingle}(g). After $T = \log_2(L/2)$ RG steps, the impurity network only contains a single tensor $M^{(T)}$, in place of the $2\times 2$ block of tensors $A^{(T)}$ that would otherwise have been in the homogeneous network $\G^{(T)}$, from which the expectation value is evaluated through the appropriate trace of $M^{(T)}$ as depicted in Fig.\ref{fig:LogFinite}.

Note that the evaluation of two-point correlators, which corresponds to contracting a network with two local impurities, can be handled in a similarly. In this case each of the impurities will transform individually in the same way as Eq.\ref{eq:Ascend} until they become adjacent to one another in the network, where they will then fuse into a single impurity after the next TNR iteration.  

\begin{figure}[!tbph]
\begin{center}
\includegraphics[width=8cm]{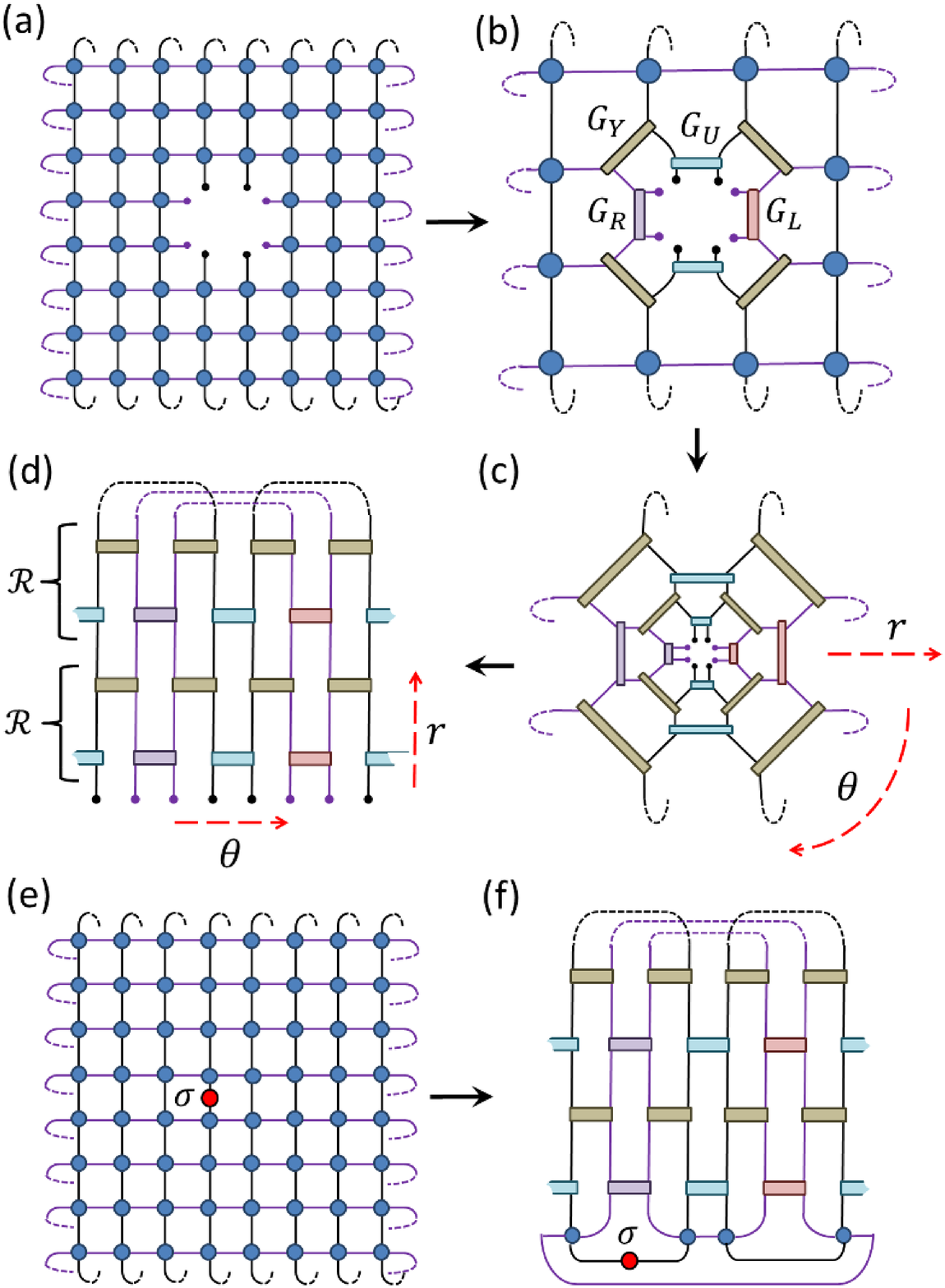}
\caption{(a) An $8\times 8$ square lattice of tensors that has had a $2\times 2$ block of tensors removed, leaving 8 open indices. (b) The lattice after one iteration of coarse-graining with TNR. (c) The lattice after a second iteration of coarse-graining with TNR. (d) The network from (c) is redrawn on the cylinder. Each double layer of the network corresponds to the ascending superoperator $\R$ as defined in Fig.\ref{fig:LogSingle}(g). (e) An $8\times 8$ square lattice of tensors with an impurity $\sigma$ positioned on a link. (f) The network from (e) after two iterations of coarse-graining with TNR.}
\label{fig:LogMany}
\end{center}
\end{figure}

The evaluation of local observables can also be formulated in a more general way, using TNR to map the network on the punctured plane to an open cylinder, analogous to the logarithmic transformation in CFT \cite{CFT1,CFT2}. This mapping using TNR was originally formulated in Ref.\onlinecite{TNRCFT}, to which we refer the interested reader for more details. A brief summary is included here for completeness. In Fig.\ref{fig:LogMany}(a) we consider a finite square lattice network that has a $2\times 2$ block of tensors removed, such that the indices connecting to the block are left open. As with the case of the impurity network considered previously, one can coarse-grain this network with TNR, performing the same projective truncations except in the immediate vicinity of the open indices, which must remain untouched. The result of this is shown in Fig.\ref{fig:LogMany}(a-d); the square lattice network [with an open `hole'] is mapped to a tensor network on a finite width cylinder. Here one end of the cylinder has free indices, which correspond to those of the open `hole', and moving along the length of the cylinder corresponds to a change of \emph{scale} in the original system, where each double row of tensors in the cylinder corresponds to an application of the ascending superoperator $\R$ as defined in Fig.\ref{fig:LogSingle}(g). 

Using the same transformations, the tensor network with local impurity, as depicted in Fig.\ref{fig:LogMany}(e), is mapped to a closed cylindrical network after coarse-graining with TNR, as depicted in Fig.\ref{fig:LogMany}(f). Contracting this network from bottom-to-top it would be equivalent to evaluating the expectation value by coarse-graining the observable as discussed previously, see Eq.\ref{eq:ImpFlow}. However, one could also evaluate the observable by contracting the network contracting from top-to-bottom, or in some other desired order. In practice, exact contraction of the network shown in Fig.\ref{fig:LogMany}(f) is computationally expensive for large bond dimension $\chi$, thus it may be necessary to use an approximate method for this evaluation. The approximate method we employ in this manuscript is based upon contracting the network layer by layer from top-to-bottom, where we approximate the boundary state as a matrix product state\cite{MPS1, MPS2} (MPS) and use the TEBD algorithm\cite{TEBD1, TEBD2} to apply each layer of gates from cylinder while maintaining an MPS representation.

\section{Benchmark results} \label{sect:Bench}
\subsection{2D classical Ising model} \label{sect:BenchClassical}
In this section we provide benchmark calculations for the binary TNR algorithm, comparing against TRG, for the partition function of $2D$ classical Ising model, as defined in Eqs.\ref{eq:Partition} and \ref{eq:HamFunct}. We begin by encoding the partition function as a tensor network as discussed in Sect.\ref{sect:TNclassical}: a four index tensor $A_{ijkl}$ is used to encode the four Boltzmann weights on the edge of a plaquette, as per Eq.\ref{eq:A}, which corresponds to having one tensor $A$ for every two spins and a tensor network with a $45$ degree tilt with respect to the spin lattice, see Fig. \ref{fig:Partition}(a-b). For convenience we then contract a $4\times 4$ square of tensors $A$ to form a new tensor $A^{(0)}$ of bond dimension $\chi=16$, which serves as the starting point for the both the TNR and TRG approaches. We apply up to 20 RG steps of either TNR or TRG, which corresponds to lattices of Ising spins with linear dimension up to $L = 4\times 2^{20} \approx 4\times 10^6$ spins. Here we define an RG step as that which maps the square lattice to a new square lattice of the same orientation, but of half the linear dimension; note that this corresponds to two steps of TRG as defined in Ref.\onlinecite{TRG}. Each of the approaches generates a sequence of coarse grained tensors, as per Eq.\ref{eq:Aflow}, where copies of tensor $A^{(s)}$ comprise the coarse-grained network after $s$ RG steps.

In the implementation of TNR we enforce reflection symmetry, as discussed in Appendix \ref{app:Reflect}, on both spatial axes, and also employ the ideas discussed in Appendix \ref{app:Reduce} to reduce the leading order cost of the TNR algorithm from scaling as $O(\chi^7)$ to scaling as $O(\chi^6)$ in terms of the bond dimension $\chi$. Furthermore we optimize tensors using the larger environment as discussed in Appendix \ref{app:Env}, and also use $Z_2$ invariant tensors [recall that the Ising model has a global $Z_2$ symmetry: it is invariant under the simultaneous flip $\sigma_k \rightarrow -\sigma_k$ of all the spins], which are employed using standard methods for incorporating global symmetries in tensor networks\cite{GlobalSym}. The TRG results have been calculated using the square-lattice TRG algorithm as presented in Ref.\onlinecite{TRG}, the cost of which also scales as $O(\chi^6)$ in terms of the bond dimension $\chi$. While the cost of TRG and TNR both scale as $O(\chi^6)$, the overall cost of a TNR calculation is greater than a TRG calculation of the same bond dimension by a constant factor, $k \approx 200$, stemming from the iterative nature of the TNR optimization. 

\begin{figure}[!t!b]
\begin{center}
\includegraphics[width=8cm]{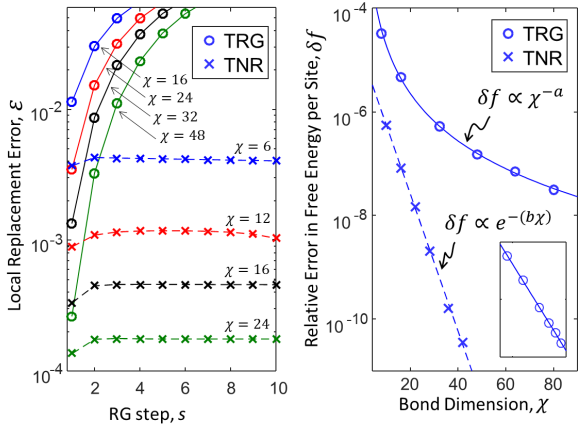}
\caption{(a) Comparison between TRG and TNR of the truncation error $\varepsilon$, as defined in Eq.\ref{eq:epsilon}, as a function of RG step $s$ in the $2D$ classical Ising model at critical temperature $T_c$. While increasing the bond dimension $\chi$ gives smaller truncation errors, the truncation errors still grow quickly as a function of RG step $s$ under TRG. Conversely, truncation errors remain stable under coarse-graining with TNR. (b) Relative error in the free energy per site ${\delta}{f}$ at the critical temperature $T_c$, comparing TRG and TNR over a range of bond dimensions $\chi$. The error from TRG is seen to diminish polynomially with bond dimension, with fit $\delta f \propto \chi^{-3.02}$ (where the inset displays the same TRG data with logarithmic scales on both axes), while the error from TNR diminishes exponentially with bond dimension, with fit $\delta f \propto \exp(-0.305 \chi)$. Extrapolation suggests that TRG would need bond dimension $\chi \approx 750$ to match the accuracy of the $\chi = 42$ TNR result. }
\label{fig:TNRerror}
\end{center}
\end{figure}

Fig. \ref{fig:TNRerror}(a) explores the truncation error $\varepsilon$, as defined in Eq.\ref{eq:epsilon}, incurred in RG step $s$ of the either the TRG or TNR approach applied to the Ising model at critical temperature, $T_c = 2/\log \left( {1 + \sqrt 2 } \right) \approx 2.269$. It is seen that, although increasing the bond dimension $\chi$ of TRG reduces the initial truncation error $\varepsilon$, the error always increases quickly as a function of RG step $s$ regardless of the bond dimension in use, a phenomenon described as the break-down of TRG at criticality in the original work by Levin and Nave\cite{TRG}. In comparison it is seen that TNR avoids such a break-down; the truncation errors $\epsilon$ remain constant over many RG steps $s$ when coarse-graining with TNR, beyond a slight increase after the initial RG step. 

\begin{figure}[!t!b]
\begin{center}
\includegraphics[width=8cm]{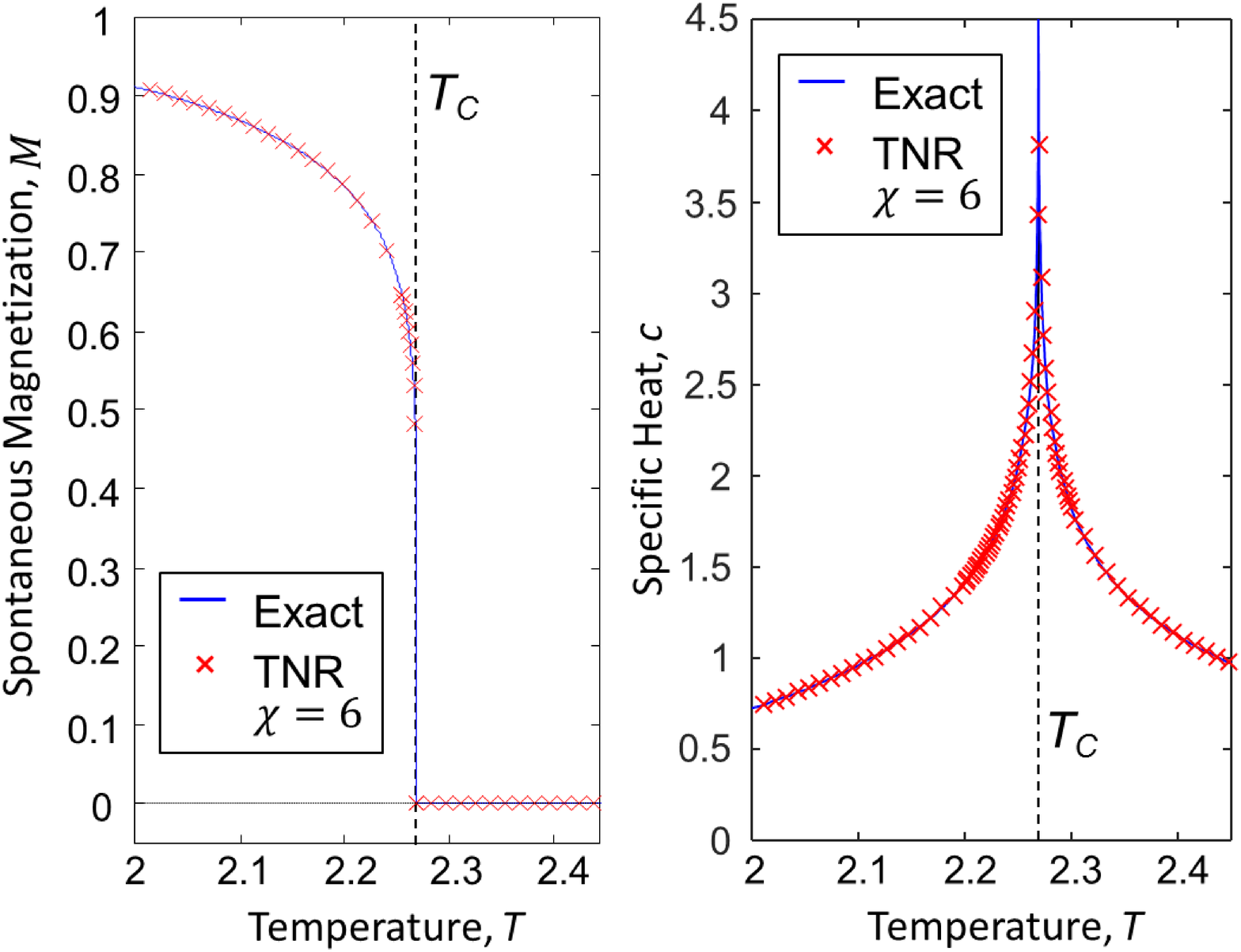}
\caption{(a) Spontaneous magnetization $M(T)$ of the $2D$ classical Ising model near critical temperature $T_c$, both exact and obtained with TNR with $\chi=6$. Even very close to the critical temperature, $T = 0.9994 \ T_c$, the magnetization $M\approx 0.48$ is reproduced to within $1\%$ accuracy. (b) Specific heat, $c(T) =  - T\frac{{{\partial ^2}f}}{{\partial {T^2}}}$, both exact and obtained using TNR with $\chi=6$.}
\label{fig:TNRphase}
\end{center}
\end{figure}

The relative error in the free energy per site, $f = -T\log(Z)/N$, at the critical temperature $T_c$ is compared for TRG and TNR over a range of bond dimensions $\chi$ in Fig.\ref{fig:TNRerror}(b). Evident is a qualitative difference in the convergence of the free energy between the approaches. In TRG the free energy $f$ is seen to converge polynomially with $\chi$, with fit $\delta f \propto \chi^{-3.02}$, while in TNR $f$ is seen to converge exponentially fast with $\chi$, with fit $\delta f \propto \exp(-0.305 \chi)$. Given that the cost of implementing the two approaches differs only by a constant factor, $k\approx 200$, for equivalent bond dimension $\chi$, it is evident that TNR can produce a significantly more accurate free energy than would be computationally viable with standard TRG. For instance, extrapolation suggests that in order to match the accuracy in the free energy for $\chi=42$ TNR, which required approximately 12 hours computation time on a laptop computer, one would need to implement TRG with $\chi\approx 750$, far beyond what is considered feasible with the approach. Figure \ref{fig:TNRphase} shows the spontaneous magnetization $M(T)$ and specific heat $c(T) =  - T\frac{{{\partial ^2}f}}{{\partial {T^2}}}$ obtained with TNR for $\chi=6$, over a range of temperatures $T$ near the critical temperature $T_c$. Remarkable agreement with the exact results is achieved throughout, even very close to the critical point $T_c$. 

\begin{figure}[!t!b]
\begin{center}
\includegraphics[width=8cm]{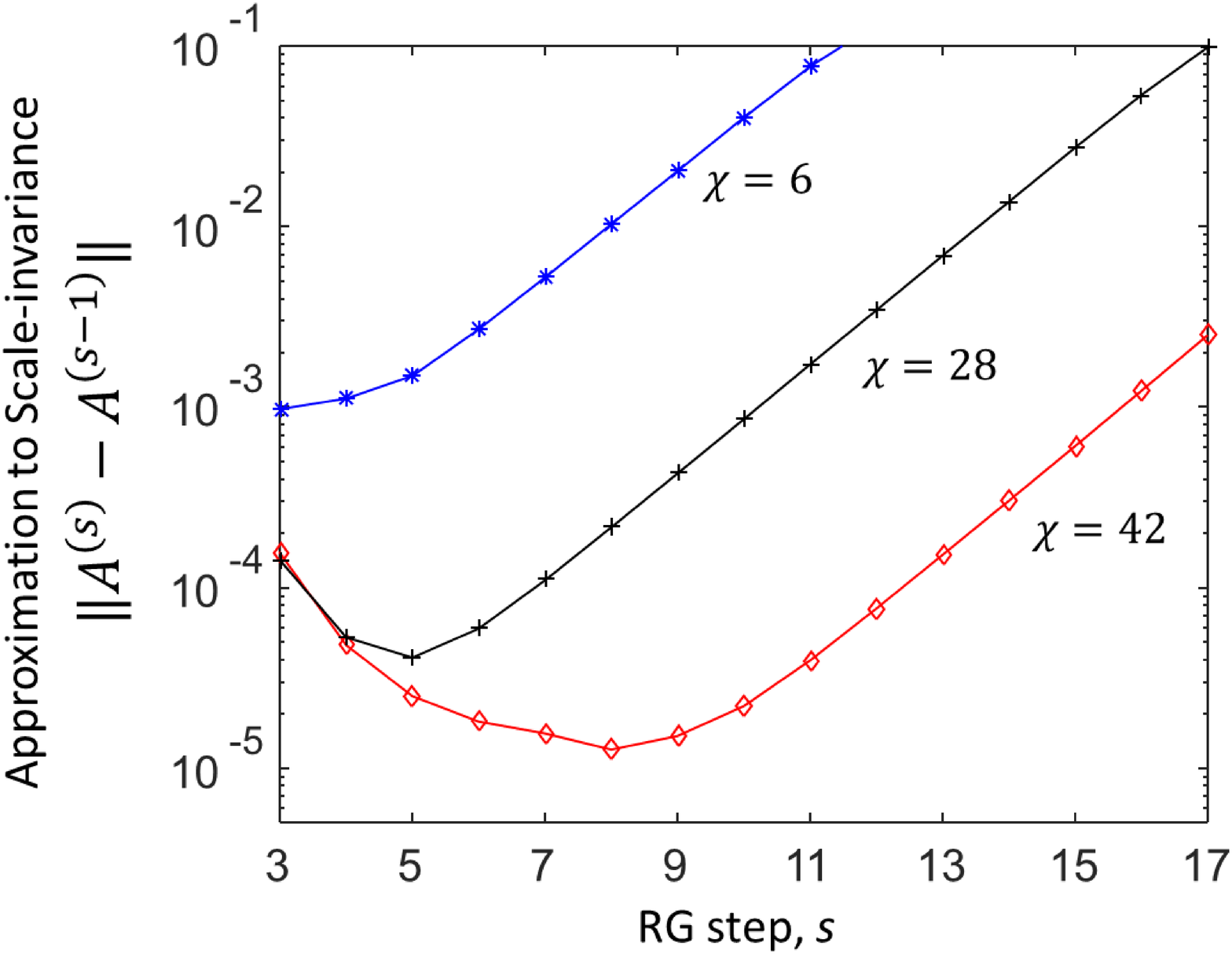}
\caption{The precision with which TNR approximates a scale-invariant fixed point tensor for the $2D$ classical Ising model at critical temperature $T_c$ is examined by comparing the difference between tensors produced by successive TNR iterations ${\delta ^{(s)}} \equiv \| {{A^{(s)}} - {A^{(s - 1)}}} \|$, where tensors have been normalized such that $\| {{A^{(s)}}} \| = 1$. The precision with which scale-invariance is approximated in the initial RG steps [small $s$] is limited by the presence of RG irrelevant terms in the lattice Hamiltonian that break scale-invariance at short-distance scales, while numerical truncation errors, which can be thought of as introducing RG relevant terms, shift the system from criticality [and thus scale invariance] in the limit of many RG steps $s$.}
\label{fig:TNRscaleinv}
\end{center}
\end{figure}

Next we explore ability of TNR to produce a scale-invariant RG flow in the space of tensors $A^{(s)}$ for the Ising model at critical temperature $T_c$. We apply up to 20 RG steps of TNR to the partition function, employing the gauge fixing strategy discussed in Appendix \ref{sect:Scale} from the third RG step onwards. The difference ${\delta ^{(s)}} \equiv \| {{A^{(s)}} - {A^{(s - 1)}}} \|$ between tensors at successive RG steps [where tensors have been normalized such that $\| {{A^{(s)}}} \| = 1$] is displayed in Fig.\ref{fig:TNRscaleinv}. For the larger $\chi$ calculations, the difference ${\delta ^{(s)}}$ reduces after the initial RG steps, before increasing again in the limit of many RG steps. This behavior is to be expected. The main limitation to realizing scale-invariance exactly in the initial RG steps is physical: the lattice system includes RG irrelevant terms that break scale-invariance at short-distance scales, but are suppressed at larger distances. On the other hand, after many RG steps the main obstruction to scale invariance is the numerical truncation errors, which can be thought of as introducing RG relevant terms, effectively shifting the flow away from criticality and thus away from scale invariance. However, use of a larger bond dimension $\chi$ reduces truncation errors, allowing TNR to not only achieve a more precise approximation to scale-invariance, but also to hold it for more RG steps.

\begin{figure}[!tbph]
\begin{center}
\includegraphics[width=8.5cm]{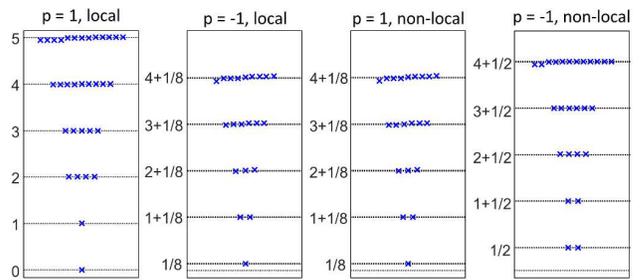}
\caption{The smallest 101 scaling dimensions of the $2D$ classical Ising at critical temperature $T_c$, obtained by diagonalizing the ascending superoperator $\R$, see Fig.\ref{fig:LogSingle}(g), using TNR with bond dimension $\chi=6$. The scaling dimensions are organized according to the parity $p=\pm 1$ (even/odd) and locality of the corresponding scaling operators. All scaling dimensions shown are reproduced to within $2\%$ of their exact values.}
\label{fig:TNRscalingdims}
\end{center}
\end{figure}

In order to demonstrate that the [approximate] fixed point map given by TNR is representative of the $2D$ Ising universality class we extract the critical data from this map. This calculation was previously carried out in Ref.\onlinecite{TNRCFT}, to which we refer the interested reader for more details. Scaling operators, with their corresponding scaling dimensions, are obtained from diagonalization of the ascending superoperator $\R$, see Fig.\ref{fig:LogSingle}(a), associated to the $s=4$ coarse-graining iteration with TNR. At this level of coarse-graining the RG irrelevant terms in the system have been sufficiently suppressed, such that a good approximation to scale-invariance is realized. The smallest 101 scaling dimensions obtained from a $\chi=6$ TNR calculation are displayed in Fig.\ref{fig:TNRscalingdims}, all of which are within $2\%$ of their exact values. Also computed were the operator product expansion (OPE) coefficients\cite{CFT1,CFT2} for the primary fields, see again Ref.\onlinecite{TNRCFT} for details, which were found to match their correct values from CFT to within $0.2\%$. These results indeed confirm that the fixed point RG map given by TNR is accurately characterizing the Ising CFT. 

\subsection{1D quantum Ising model} \label{sect:BenchQuantum}
In this section we provide benchmark calculations for the binary TNR algorithm applied to the Euclidean path integral of the $1D$ quantum Ising model, which has Hamiltonian $H_\textrm{Is.}$ defined,
\begin{equation} \label{eq:HamIs}
H_\textrm{Is.} = \sum\limits_k {\left( {{\sigma_k^x}{\sigma_{k + 1}^x} + \lambda {\sigma_k^z}} \right)} 
\end{equation}
where $\sigma^x$ and $\sigma^z$ are Pauli matrices, and $\lambda$ represents the magnetic field strength. For convenience we perform an initial blocking step, in which blocks of four spins are combined into effective sites of local dimension $d = 16$, and then generate the tensor network representation of the Euclidean path integral as explain in Sect.\ref{sect:TNquantum}. We typically use a time step of $\tau = 0.002$ and coarse-grain in the Euclidean time direction using 10 iterations of the single dimension coarse-graining scheme explained in Appendix \ref{app:Compress}, before beginning the TNR calculation. 

\begin{figure}[!t!b]
\begin{center}
\includegraphics[width=8.5cm]{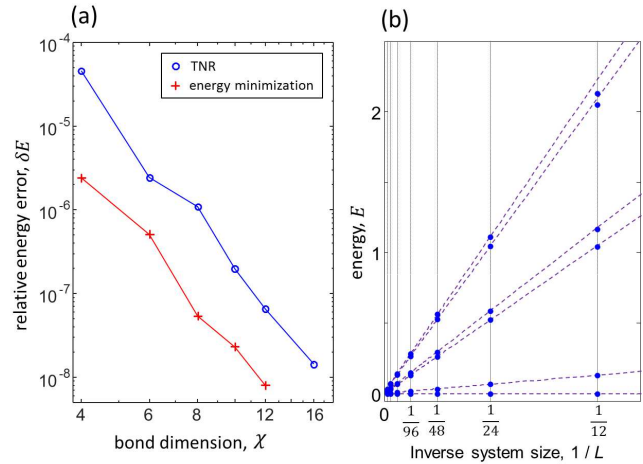}
\caption{(a) Relative error in the energy of scale-invariant MERA optimized for the ground state of the $1D$ quantum Ising model at criticality in terms of bond dimension $\chi$, comparing MERA optimized using TNR to those optimized using variational energy minimization. Energy minimization produces MERA with a more accurate approximation to the ground energy, but is significantly more computationally expensive [with a computational cost that scales as $O(\chi^9)$ versus as $O(\chi^6)$ for TNR]. (b) Low energy eigenvalues of the $1D$ quantum Ising model at criticality as a function of 1/L, computed with $\chi=12$ TNR. Discontinuous lines correspond to the finite-size CFT prediction, which ignores corrections of order $L^{-2}$.}
\label{fig:EnergyPlot}
\end{center}
\end{figure}

To start with, we compare the performance of the TNR algorithm as a means to optimize a MERA\cite{MERA} versus the energy minimization approach of Ref.\onlinecite{Algorithms}. The ground state of the Ising Hamiltonian $H_\textrm{Is.} $ at critical magnetic field strength, $\lambda = 1$, is represented using a scale-invariant MERA\cite{Algorithms, criticality2, criticality3, criticality4, criticality5}, here consisting of three transitional layers followed by infinitely many copies of a scale-invariant layer, obtained in two different ways. In the first approach, we apply four coarse-graining iterations of TNR to the Euclidean path integral of $H_\textrm{Is.} $, after which the flow in the space of tensors has reached an approximate scale-invariant fixed-point. A scale-invariant MERA is then built from certain tensors that were produced from the TNR calculation, specifically the disentanglers $u$ and isometries $w$, as described in the Ref.\onlinecite{TNRtoMERA}. Here the first three iterations of TNR generate the three transitional layers, while the fourth iteration of TNR is used to build the scale-invariant layer. In the second approach, we use the energy minimization algorithm of Refs.\onlinecite{Algorithms,criticality4} to directly optimize a scale-invariant MERA [with three transitional layers preceding the scale-invariant layers] by iteratively minimizing the expectation value of the $H$. Presented in Fig. \ref{fig:EnergyPlot}(a) is the comparison between the ground energy error $\delta E$ obtained by the two methods. For an equivalent bond dimension $\chi$, energy minimization produces a MERA with smaller error $\delta E$ than TNR, by a factor $k \approx 10$ that is roughly independent of $\chi$. However, optimization using TNR is also computationally cheaper than optimization using the energy minimization algorithm; the cost of TNR scales as $O(\chi^6)$ in terms of bond dimension $\chi$, while the energy minimization scales as $O(\chi^9)$. In addition, the energy minimization algorithm, which iteratively sweeps over all MERA layers, requires more iterations to converge than does TNR. Taking these considerations into account, TNR can be seen to be the more efficient approach to obtain a ground state MERA to within a given level of accuracy in the energy $\delta E$, see Ref.\onlinecite{TNRtoMERA} for additional details. Thus the TNR approach represents a useful, alternative means for optimizing a MERA and, by extension, is promising as a tool for the exploration of ground states of quantum systems.

In Fig.\ref{fig:EnergyPlot}(b) the low energy excitation spectra of $H_\textrm{Is.}$ at criticality, $\lambda = 1$, is plotted as a function of system size $L$. This was computed using TNR to coarse-grain the Hamiltonian $H_\textrm{Is.}$, as explained in Ref.\onlinecite{TNRtoMERA}, which was then diagonalized on a finite lattice with periodic boundaries. The excitation spectra reproduce the expected $1/L$ scaling with system size and match the predictions from CFT\cite{CFT1, CFT2}, demonstrating that, in addition to the ground state, TNR can accurately approximate the low-energy eigenstates of $H_\textrm{Is.}$.

\begin{figure}[!t!b]
\begin{center}
\includegraphics[width=8.5cm]{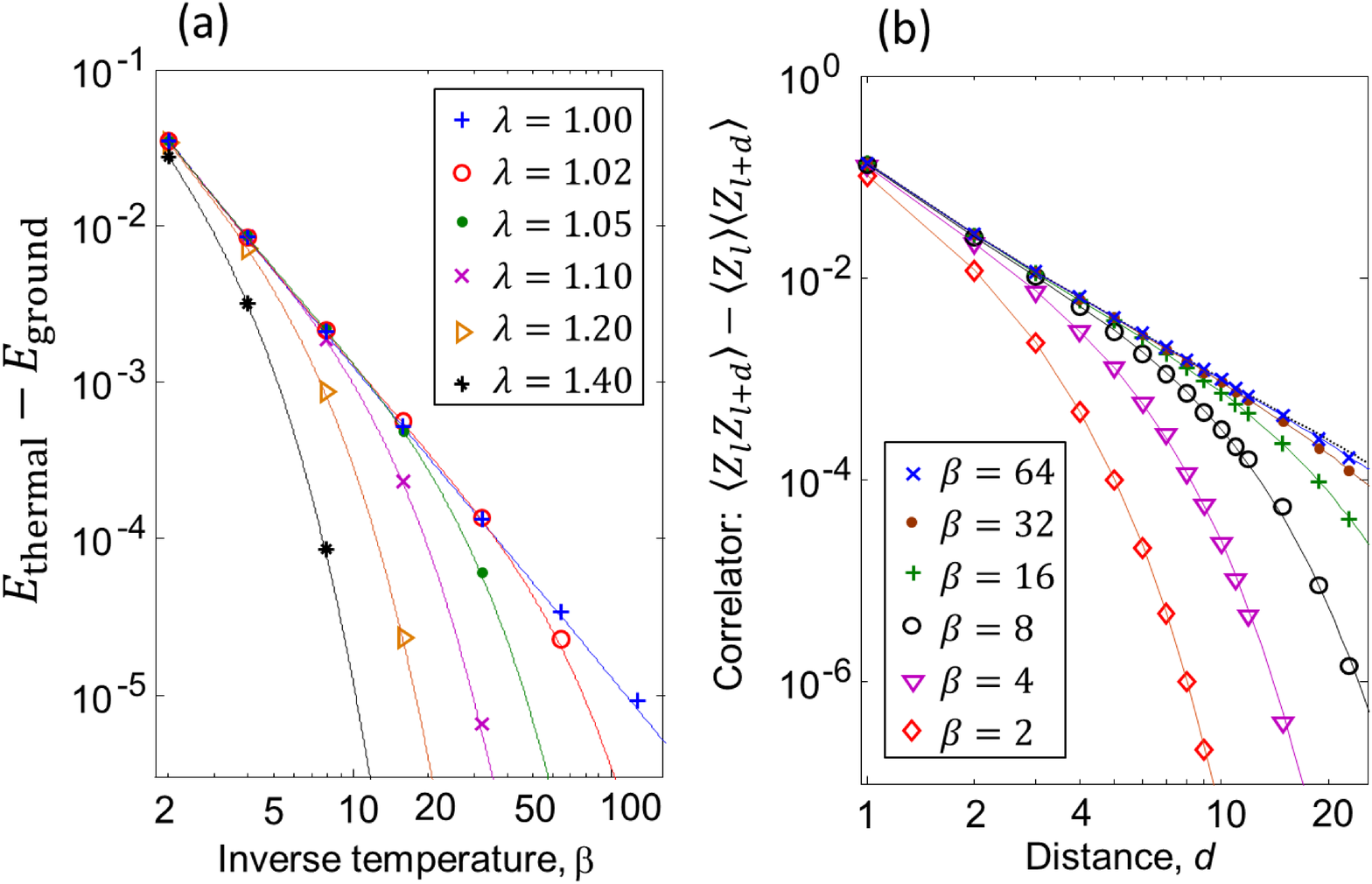}
\caption{(a) Thermal energy per site (above the ground state energy) as a function of the inverse temperature $\beta$, for the $1D$ quantum Ising model in an infinite chain, for different values of magnetic field $\lambda$. Data points are computed with $\chi=12$ TNR while continuous lines correspond to the exact solution. (b) Connected two-point correlators at the critical magnetic field $\lambda = 1$, as a function of the distance $d$, for several values of $\beta$. Data points are computed with $\chi=12$ TNR while continuous lines again correspond to the exact solution.}
\label{fig:ThermalPlot}
\end{center}
\end{figure}

Next we explore the use of TNR for computing expectation values of finite temperature thermal states. This is achieved by applying TNR to the tensor network corresponding to $e^{-\beta H_\textrm{Is.}}$ for inverse temperature $\beta$, as discussed in Sect.\ref{sect:TNquantum}, which is then used to generate a thermal state MERA as explained in Ref.\onlinecite{TNRtoMERA}. The thermal energy per site as a function of $\beta$ is displayed in Fig.\ref{fig:ThermalPlot}(a) for several different magnetic field strengths $\lambda$. In the gapped regime, $\lambda > 1$, the thermal energy converges exponentially quickly to the ground energy in the limit of large $\beta$ [or small temperature $T$], while at the gapless critical point, $\lambda = 1$, the thermal energy converges polynomially quickly to the ground energy with $\beta$. The results from $\chi = 12$ TNR accurately reproduce the exact energies over the range of parameters considered. Two-point correlation functions of the critical, $\lambda = 1$, system are examined in Fig.\ref{fig:ThermalPlot}(b) over a range of inverse temperatures $\beta$, where correlations are seen to decay faster at smaller $\beta$ [or equivalently at higher temperature $T$] due to thermal fluctuations. Again, the results from $\chi = 12$ TNR accurately reproduce the exact correlations, indicating that TNR can be used to approximate thermal states of quantum systems over a wide range of temperatures.

\section{Discussion} \label{sect:Discussion}
After reviewing the conceptual foundations for real-space renormalization of partition functions and Euclidean path integrals, when expressed as tensor networks, a self-contained description of the algorithm for employing TNR to study properties of classical and quantum many-body systems was provided. 

Benchmark results, provided in Sect.\ref{sect:Bench}, demonstrated some of the advantages of TNR. These include (i) providing a computationally sustainably coarse-graining transformation even for systems at or near critical point [i.e. one that can be iterated many times without increase in truncation error], (ii) convergence in the RG flow of tensors to a scale-invariant fixed point for a critical system [which then allows calculation of the critical data from the fixed point RG map directly], (iii) providing an alternate, more efficient means for optimizing a MERA for the ground state of a quantum system, and (iv) providing a means to accurately study properties of thermal states of quantum systems over a wide range of temperatures $T$. With regards to (i) above, TNR was demonstrated to overcome a major obstacle of previous schemes for renormalization of tensor networks, such as TRG, which exhibit a computational breakdown when near or at a critical point. As a consequence, the free energy per site $f$ of the $2D$ classical Ising model at criticality was seen to converge to the exact value exponentially faster in bond dimension $\chi$ with TNR than with the previous TRG approach.

Future work shall include the development and implementation of TNR schemes for the coarse-graining of networks on $3D$ lattices, which could be applied to study $3D$ classical statistical and $2D$ quantum many-body systems. A TNR scheme for $3D$ lattices would offer an alternative, potentially more efficient, means to optimize a $2D$ MERA over the previous [often prohibitively expensive] strategies based upon energy minimization\cite{2DMERA2, 2DMERA1}. The TNR approach presented in this manuscript can also be used to compute the norm $\bk{\Psi | \Psi}$ of a $2D$ quantum many-body state encoded in a PEPS \cite{PEPS1, PEPS2}, thus it could be incorporated as a key part of an algorithm for simulation of $2D$ quantum many-body systems using PEPS\cite{PEPS3}. 

The author thanks Markus Hauru and Guifre Vidal for insightful comments. The author acknowledges support by the Sherman-Fairchild Foundation and by the Simons Foundation (Many Electron Collaboration).

\appendix
\section{Coarse-graining along a single dimension} \label{app:Compress}
In this appendix we describe a scheme that coarse-grains a square-lattice along one dimension only, which is similar in effect to the higher order tensor renormalization group (HOTRG) method introduced in Ref.\onlinecite{TRG3D}. It is useful to perform this coarse-graining to rescale one dimension before applying the TNR algorithm if the initial tensor network is highly anisotropic in the strength of its correlations [which occurs, for instance, when the network represents the Euclidean path integral of a quantum system expanded with a very small time-step $\tau$]. The preliminary coarse-graining can generate a network that is more isotropic in the strength of its correlations and thus more suitable as a starting point for the TNR approach, which then rescales both lattice dimensions equally.

\begin{figure}[!t!b]
\begin{center}
\includegraphics[width=8cm]{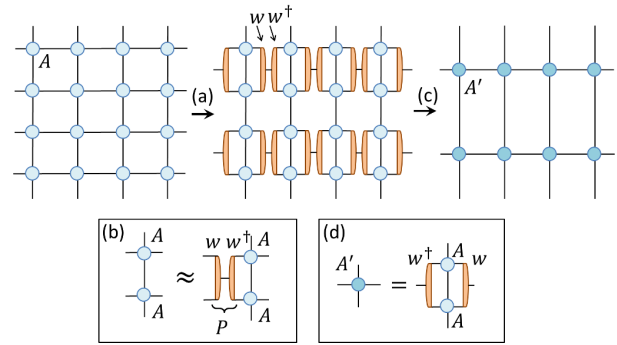}
\caption{An iteration of a coarse-graining scheme that acts to compress only the vertical dimension of the square-lattice tensor network. (a) A projective truncation, involving isometries $w$, is employed. (b) Details of the projective truncation from (a); here a projector $P$, comprised of an isometry $w$ and its conjugate, is applied to a pair of $A$ tensors. (c) The coarse-grained network, comprised of tensors $A'$ is given via tensor contractions. (d) Definition of the coarse-grained tensor $A'$.}
\label{fig:TimeCompression}
\end{center}
\end{figure}

An iteration of the coarse-graining, which acts to compress the vertical dimension of the square-lattice tensor network by a factor of two, is depicted in Fig.\ref{fig:TimeCompression}. The first step, as shown Fig.\ref{fig:TimeCompression}(a), involves a projective truncation that acts upon a pair of $A$ tensors, as further detailed in Fig.\ref{fig:TimeCompression}(b). The isometries $w$ that comprise the projector can be optimized in the standard way, as discussed in Sect.\ref{sect:Project}. The second step, as shown in Fig.\ref{fig:TimeCompression}(c), involves contraction of a pair of $A$ tensors with isometries $w$, as detailed in Fig.\ref{fig:TimeCompression}(d), to generate the tensors $A'$ of the coarse-grained network.

\section{Ternary TNR scheme} \label{app:Ternary}
The binary TNR scheme presented in Sect.\ref{sect:binTNR} is one of many possible TNR schemes for the square-lattice; here we describe a ternary TNR scheme which reduces the linear dimension of the lattice by a factor of 3 with each iteration [or, equivalently, coarse-grains a $3\times 3$ block of tensors into a new tensor with each iteration]. One advantage of the ternary TNR scheme is that single tensor impurities can remain single tensor impurities under coarse-graining, in contrast to the binary TNR scheme in which impurities spread to $2\times 2$ blocks as depicted Fig.\ref{fig:LogSingle}. This property could be convenient for certain types of calculation, such as in the evaluation of conformal data from a critical system. Through a similar derivation as presented in Ref.\onlinecite{TNRtoMERA}, where the binary TNR scheme was shown to yield a binary MERA, the ternary TNR scheme can be shown to yield a ternary MERA\cite{Algorithms}. 

\begin{figure}[!t!b]
\begin{center}
\includegraphics[width=8cm]{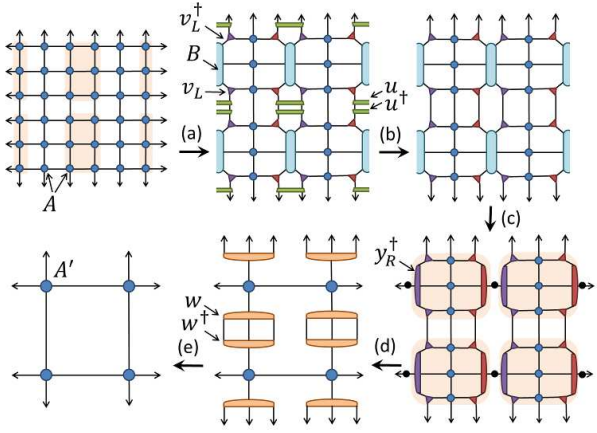}
\caption{A depicted of an iteration of the ternary TNR scheme, which maps a square-lattice tensor network to a coarse-grained square-lattice of one-third the linear dimension of the original. (a) A projective truncation, detailed in Fig.\ref{fig:TernSchemeB}(a). (b) Pairs of disentanglers $u$ annihilate to identity. (c) A projective truncation, detailed in Fig.\ref{fig:TernSchemeB}(b). (d) A projective truncation, detailed in Fig.\ref{fig:TernSchemeB}(c). (e) Pairs of isometries $w$ annihilate to identity, yielding the coarse-grained square lattice network.}
\label{fig:TernSchemeA}
\end{center}
\end{figure}

\begin{figure}[!t!b]
\begin{center}
\includegraphics[width=8cm]{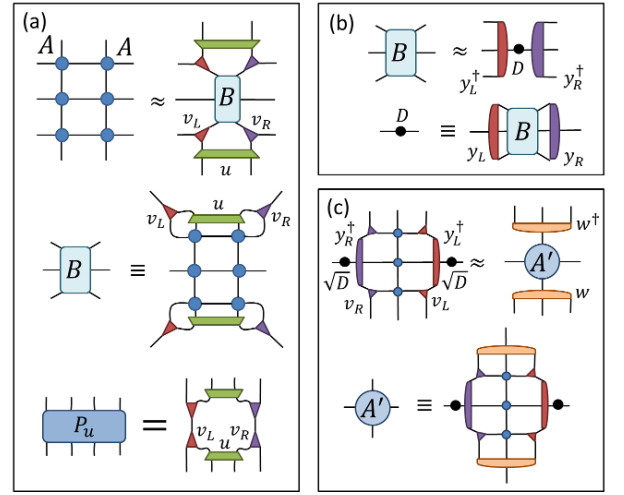}
\caption{Overview of the projective truncations involved in the ternary TNR scheme. (a) A projective truncation, implemented by projector $P_u$ composed of isometries $v_L$, $v_R$ and disentangler $u$, acts upon a $3\times 2$ block of $A$ tensors. (b) A projective truncation, formed from isometries $y_L$, $y_R$ and their conjugates, acts upon $B$ tensors. (c) A projective truncation, formed from isometry $w$ and its conjugate, acts to give the new four index tensor $A'$ of the coarse-grained network.  }
\label{fig:TernSchemeB}
\end{center}
\end{figure}

The steps of an iteration of the ternary TNR scheme are shown in Fig.\ref{fig:TernSchemeA}(a-e). These steps are as follows: (a) a projective truncation involving isometries $v_L$, $v_R$ and disentangler $u$ as detailed in Fig.\ref{fig:TernSchemeB}(a), (b) annihilation of conjugate pairs of disentanglers $u$ to identity, (c) a projective truncation involving isometries $y_L$ and $y_R$ as detailed in Fig.\ref{fig:TernSchemeB}(b), (d) a projective truncation involving isometry $w$ as detailed in Fig.\ref{fig:TernSchemeB}(c), (e) annihilation of conjugate pairs of the isometry $w$ to identity. The four-index tensor $A'$ of the coarse-grained network, as defined in Fig.\ref{fig:TernSchemeB}(c), now accounts for a $3\times 3$ block of tensors $A$ from the original square-lattice network. 

\section{Isotropic TNR scheme} \label{app:Isotropic}
In this appendix we present another TNR scheme for the square-lattice. Like the binary TNR scheme presented in the main text, this scheme also reduces the linear dimension of the lattice by a factor of 2 with each iteration. However, unlike the binary TNR scheme, this scheme treats both dimensions of the lattice equally, for which we refer to it as the \emph{isotropic} TNR scheme. Thus, if the starting network that is invariant under $90^\circ$ rotations, such as that corresponding to the partition function of an isotropic $2D$ classical statistical model, this TNR scheme can preserve the rotational symmetry under coarse-graining.

\begin{figure}[!t!b]
\begin{center}
\includegraphics[width=8.5cm]{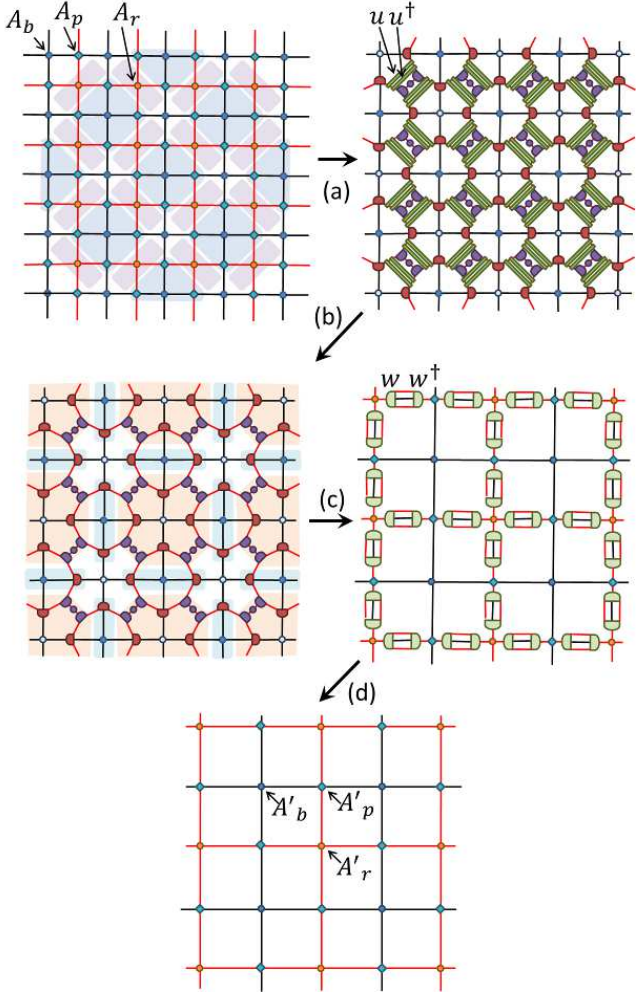}
\caption{Depiction of a single iteration of the isotropic TNR scheme, which maps a square-lattice network, comprised of tensors $A_b$, $A_p$ and $A_r$, to a coarser lattice of the same type. (a) Two types of projective truncation are made, as detailed in Fig.\ref{fig:IsoSchemeB}(a-b). (b) Conjugate pairs of disentanglers $u$ annihilate to identity. (c) Two types of projective truncation are made, as detailed in Fig.\ref{fig:IsoSchemeB}(c-d). (d) Conjugate pairs of isometries $w$ annihilate to identity, yielding a coarse-grained network comprised of tensors ${A'}_b$, ${A'}_p$ and ${A'}_r$.}
\label{fig:IsoSchemeA}
\end{center}
\end{figure}

The isotropic TNR scheme is applied to a square-lattice network with a four-site unit cell, where it is assumed that the network consists of three types of four index tensor, $A_b$, $A_p$ and $A_r$. As shown in Fig.\ref{fig:IsoSchemeA} the lattice has a $2\times 2$ unit cell where each $A_b$ tensor connects with four $A_p$ tensors in the network, and likewise each $A_r$ tensor also connects with four $A_p$. It is assumed that $A_p$ and $A_r$ tensors are invariant with respect to $90^\circ$ rotations, while $A_p$ tensors are invariant with respect to $180^\circ$ rotations; under this assumption the square-network itself is invariant with respect to $90^\circ$ rotations centered about either an $A_p$ or $A_r$ tensor. Notice that a uniform [one-site unit cell] square lattice is just a special case of this four-site unit cell lattice, where indices on one of the sub-lattices, such as indices connected to $A_r$ tensors, are fixed at trivial dimension, i.e. bond dimension $\chi = 1$. Thus this isotropic TNR scheme can be directly applied to the partition function of an isotropic $2D$ classical statistical model, when it is represented as a uniform square-lattice network of tensors $A$ that are invariant with respect to $90^\circ$ rotations.

\begin{figure}[!t!b]
\begin{center}
\includegraphics[width=8.5cm]{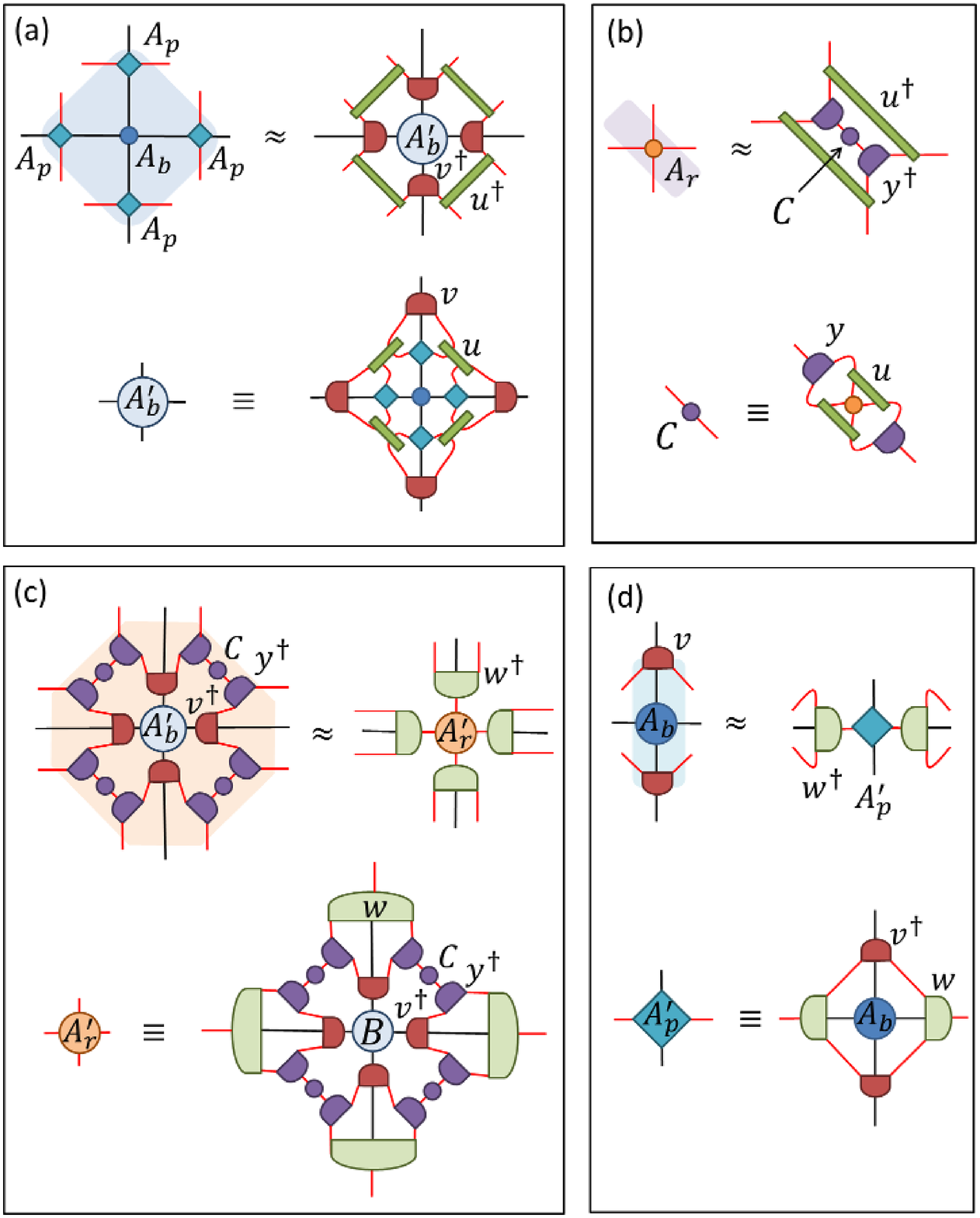}
\caption{Overview of the projective truncations involved in an iteration of the isotropic TNR scheme. (a) A projective truncation, involving isometries $v$ and disentanglers $u$, is enacted upon a product of four $A_p$ tensors and a $A_b$ tensor. The coarse-grained tensor ${A'}_b$ is also defined. (b) A projective truncation, involving isometries $y$ and disentanglers $u$, is enacted upon $A_r$ tensors. (c) A projective truncation, involving isometries $w$, is enacted, yielding coarse-grained tensors ${A'}_r$. (d) A final projective truncation, again involving isometries $w$, is applied to a product of $A_b$ and $v$ tensors, yielding coarse-grained tensors ${A'}_p$.}
\label{fig:IsoSchemeB}
\end{center}
\end{figure}

The steps of a coarse-graining iteration of the isotropic TNR scheme are shown in Fig.\ref{fig:IsoSchemeA}(a-d). The initial step, shown in Fig.\ref{fig:IsoSchemeA}(a), involves two different projective truncations. One of the projective truncations, as detailed in Fig.\ref{fig:IsoSchemeB}(a), acts upon a block of four $A_p$ tensors and a $A_b$ tensor, while the other acts upon individual $A_r$ tensors as detailed in Fig.\ref{fig:IsoSchemeB}(b). In the second step of the iteration, shown in Fig.\ref{fig:IsoSchemeA}(b), conjugate pairs of disentanglers $u$ annihilate to the identity. A pair of projective truncations are used in the third step of the iteration, seen in Fig.\ref{fig:IsoSchemeA}(c). These projective truncations, which involve a projector composed of an isometry $w$ and its conjugate, are detailed in Fig.\ref{fig:IsoSchemeB}(c-d). In the final step of the iteration, as shown in Fig.\ref{fig:IsoSchemeA}(d), certain isometries $w$ annihilate to identity with their conjugates, yielding the coarse-grained network. Notice that the tensors ${A_b}'$, ${A_r}'$ and ${A_p}'$ of the coarse grained network, as defined in Fig.\ref{fig:IsoSchemeB}, possess the same rotational symmetry as the corresponding initial tensors $A_b$, $A_r$ and $A_p$, and also that the coarse-grained network has the same unit cell [in terms of the coarse-grained tensors] as the initial network, but is reduced by a factor of 2 in linear dimension.  

\section{Reflection symmetry} \label{app:Reflect}
In this appendix we describe the details, given a tensor network $\G$ that is symmetric with respect to spatial reflections [perhaps in conjunction with complex conjugation] along one axis, for how the symmetry can be preserved under coarse-graining with TNR. In the case that $\G$ represents a Euclidean path integral the presence of this symmetry follows from the Hermicity of the Hamiltonian, and is thus always present, while in the case that $\G$ represents a partition function the symmetry is present if the underlying $2D$ classical statistical model is invariant under spatial reflection along an axis. Proper exploitation of this symmetry is desirable as it can significantly simplify the TNR algorithm.

\begin{figure}[!t!b]
\begin{center}
\includegraphics[width=8cm]{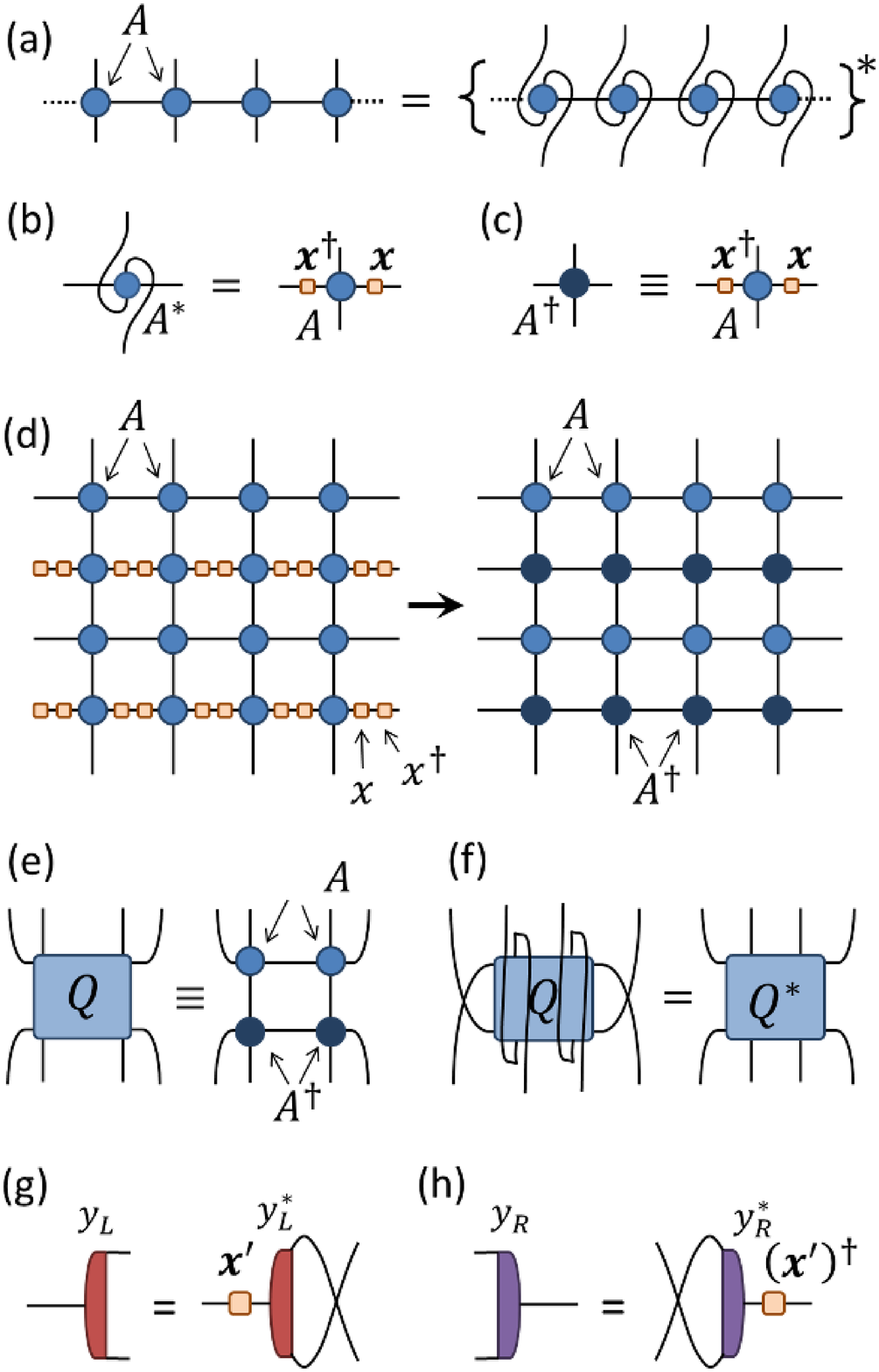}
\caption{(a) The tensor $A$ is Hermitian symmetric if a row of such tensors is invariant with respect to permutation of top-bottom indices in conjunction with complex conjugation. (b) Permutation and complex conjugation of Hermitian symmetric $A$ is equivalent to a gauge change, enacted by some unitary matrix $\bf{x}$, on its horizontal indices. (c) Definition of $A^{\dag}$. (d) A gauge change is performed on the tensors of every second row of the square-lattice network. (e) Definition of tensor $Q$. (f) Tensor $Q$ is seen to be Hermitian. (g-h) Constraints for isometries $y_L$ and $y_R$ necessary such that the tensors of the coarse-grained network remain Hermitian symmetric.}
\label{fig:GaugeRef}
\end{center}
\end{figure}

We say the homogeneous tensor network $\G$ is Hermitian symmetric with respect to the horizontal axis if a row of tensors, which form a matrix product operator (MPO), is invariant with respect to permutation of top-bottom indices in conjunction with complex conjugation, see also Fig.\ref{fig:GaugeRef}. If a row of tensors satisfies the above definition of reflection symmetry, then it can be shown that permutation of top-bottom indices and complex conjugation of tensor $A$ is equivalent to enacting a unitary gauge change on its horizontal indices,
\begin{equation} \label{eq:Gauge}
A^{\dag} \equiv {\left( {{A_{ilkj}}} \right)^ * } = \sum\limits_{j',l'} {{\bf{x}}_{j,j'}^ * {A_{ijkl}}{{\bf{x}}_{l',l}}},   
\end{equation}
see also Fig.\ref{fig:GaugeRef}(b-c). Here the indices of tensor $A_{ijkl}$ are labeled clockwise from the top, as per Fig.\ref{fig:Partition}(a), and $\bf{x}$ is some unitary matrix. 

In the first step of the TNR iteration, as discussed in Sect.\ref{sect:binTNR}, it is useful to enact the unitary gauge change $\bf{x}$ on every second row of tensors in the network $\G$ as depicted in Fig.\ref{fig:PathIntegral}(b). Let us define tensor $Q$ as the tensor formed from contracting two copies of $A$ and two copies of $A^\dag$ together as depicted in Fig.\ref{fig:PathIntegral}(d). The reason that the initial gauge transformation on $\G$ is useful is that it allows tensor $Q$ to be Hermitian under exchange of top-bottom indices, thus at the first step of the TNR iteration the same projector $P_u$ may be used both on the top and bottom of $Q$, see Fig.\ref{fig:TNRsteps}(a-c). The reflection symmetry can be preserved under the TNR iteration, such that the tensors $A'$ of the coarse grained network satisfy Eq.\ref{eq:Gauge} for some unitary matrix ${\bf{x}}'$, if the isometries $y_L$ and $y_R$ used in the second step of the TNR iteration satisfy the relation shown in Fig.\ref{fig:GaugeRef}(e-f). This relation states that the isometries should be invariant under complex conjugation in conjunction with permutation of their incoming indices and a unitary gauge change, enacted by unitary matrix ${\bf{x}'}$, on their outgoing index. Isometries $y_L$ and $y_R$ satisfying this relation can be obtained by symmetrizing the environments $\Gamma_{y_L}$ and $\Gamma_{y_R}$ of the isometries during their optimization, similar to previously strategies for preserving reflection symmetry in MERA described in Ref.\onlinecite{ImpurityMERA1}.

Finally, we remark these ideas can be extended such that reflection symmetry along the vertical axis can be preserved [simultaneously with that on the horizontal axis], if both symmetries are present in the initial network.  

\section{Reduction of computational cost} \label{app:Reduce}
\begin{figure}[!t!b]
\begin{center}
\includegraphics[width=8.5cm]{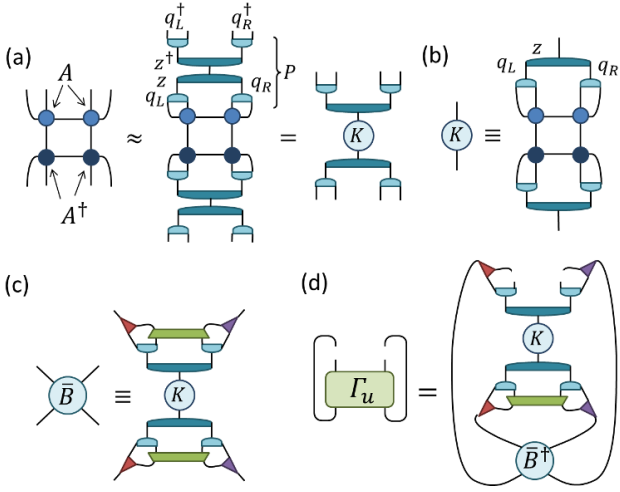}
\caption{(a) Details of an optional, preliminary projective truncation for the binary TNR scheme where two copies of a projector $P$, which is formed from a product isometric tensors $q_L$, $q_R$, $z$ and their conjugates, are enacted upon a $2\times 2$ block of $A$ tensors. (b) Definition of tensor $K$. (c) Definition of tensor $\bar{B}$, which differs from the previous $B$, see Fig.\ref{fig:TNRsteps}(b), only by an amount related to the truncation error of $P$ in (a). (d) Detail of the environment $\Gamma_u$ computed from $\bar{B}$. The cost of contracting this network sales as $O(\chi^6)$ in terms of the bond dimension $\chi$, as opposed to $O(\chi^7)$ for the previous environment of Fig.\ref{fig:Optimization}(c).}
\label{fig:CostReduce}
\end{center}
\end{figure}

The cost of the TNR algorithm as described in Sect.\ref{sect:TNR} scales as $O(\chi^7)$ in terms of the bond dimension $\chi$; in this appendix here we describe how this cost scaling can be reduced to $O(\chi^6)$. This reduction in cost is achieved by doing an additional projection truncation at the start of each TNR iteration; specifically this projective truncation is enacted on $2\times 2$ blocks of $A$ tensors, with two of the tensors conjugated as discussed in Appendix \ref{app:Reflect}, before the projective truncation step of Fig.\ref{fig:TNRsteps}(a-c). The projector $P$ involved in this step is composed of isometries $q_l$, $q_r$ and $z$ [and their complex conjugates], as depicted in Fig.\ref{fig:CostReduce}, and can be optimized with the standard iterative SVD approach as described in Sect.\ref{sect:Project}.

After this initial projective truncation the cost of the subsequent step of the TNR iteration is reduced. Fig.\ref{fig:CostReduce}(c) depicts the new tensor $\bar{B}$, which differs from the tensor $B$ in Fig.\ref{fig:TNRsteps}(b) only by an amount related to the truncation error $\varepsilon$ of the initial projection step, but can be computed with a cost that scales $O(\chi^6)$, as opposed to $O(\chi^7)$ for computing $B$. Likewise the environment $\Gamma_u$ of disentangler $u$ when expressed in terms of $\bar B$, see Fig.\ref{fig:CostReduce}(d), can also be computed with cost $O(\chi^6)$ instead of the cost $O(\chi^7)$ associated to computing the previous environment of Fig.\ref{fig:Optimization}(c). Similarly, the environments of isometries $v_L$ and $v_R$ can also be computed with cost $O(\chi^6)$ when using $\bar B$. Thus, when using the results of this appendix, no operation required to implement the binary TNR scheme has a cost scaling of greater than $O(\chi^6)$. 

\section{Optimization using a larger environment} \label{app:Env}
In this appendix we discuss how the accuracy of the binary TNR scheme, for given bond dimension $\chi$, can be improved by taking a larger environment into account at each truncation step. This follows similar ideas introduced in Ref.\onlinecite{TRGenv} to improve TRG by taking the local environment into consideration at each truncation step.

\begin{figure}[!t!b]
\begin{center}
\includegraphics[width=8cm]{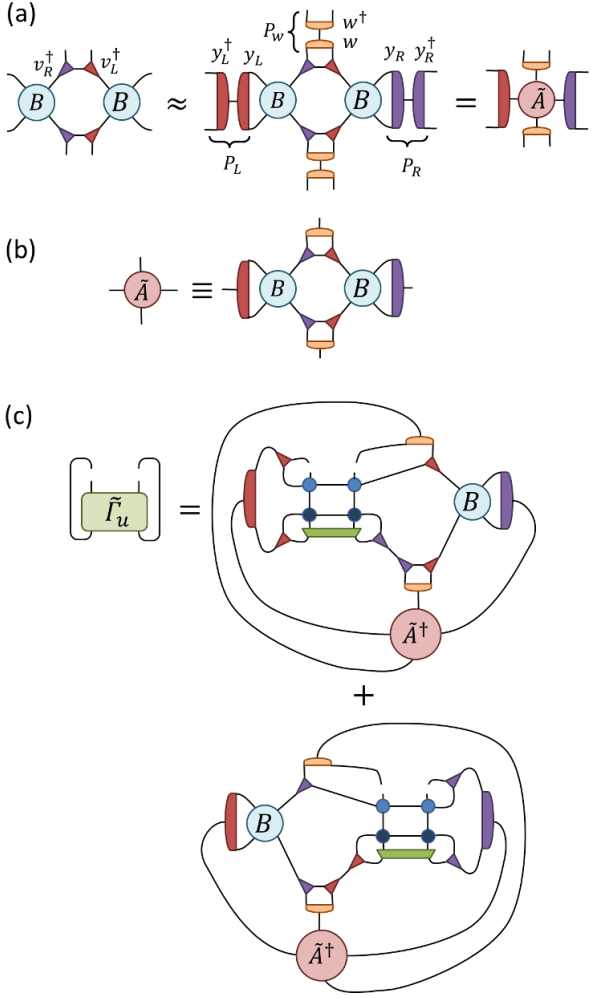}
\caption{(a) Isometric tensors $y_L$, $y_R$, $w$ yielding a more accurate coarse-graining transformation can found by optimizing them to minimize the truncation error when applied to a region of the network, here containing two copies of the $B$ tensor, that is larger than was previously considered in Fig.\ref{fig:TNRsteps}(d,f). Notice that minimizing the truncation error is equivalent to maximizing the norm of $\tilde A$. (b) Definition of tensor $\tilde A$. (c) The two contributions to the modified environment $\tilde{\Gamma}_u$ for disentangler $u$, as generated from $\bnorm{\tilde A}^2$.}
\label{fig:BigEnv}
\end{center}
\end{figure}

The TNR approach is based upon the use \emph{projective truncations} to implement coarse-graining transformations, as discussed in Sect.\ref{sect:Project}. A projective truncation involves application of a projector $P$ to a local sub-network of tensors $\F$, where it is desired that $P$ acts on $\F$ as an approximate resolution of the identity, see Eq.\ref{eq:proj4}. Use of a larger sub-network $\F$ typically allows a more accurate truncation, as the projector $P$ can take into account correlations from a larger region of the network. Fig.\ref{fig:BigEnv}(a) depicts a larger sub-network, consisting of two copies of the $B$ tensor in addition to $v_L$ and $v_R$ tensors, that can be used in the determination of the projectors $P_L$, $P_R$ and $P_w$ in the second and third step of the TNR iteration. The condition that projectors $P_L$, $P_R$ and $P_w$ act with small truncation error on this sub-network, as shown in Fig.\ref{fig:BigEnv}(a), is less restrictive than the condition previously imposed on the projectors in Sect.\ref{sect:TNR} [which used a smaller sub-network], thus potentially allows for more accurate projectors to be chosen. The isometric tensors $\{ y_L, y_R, w\}$ that compose these projectors can be chosen to minimize the truncation error $\varepsilon$ by optimizing them to maximize ${\bnorm{\tilde{A}}^2}$, with tensor $\tilde A$ as defined in Fig.\ref{fig:BigEnv}(b), which effectively replaces the two separate optimizations depicted previously in Fig.\ref{fig:TNRsteps}(d,f).

We would also like to use the larger sub-network in the optimization of the disentangler $u$. Given that the $B$ tensors depend on the disentanglers $u$, as depicted in Fig.\ref{fig:TNRsteps}(b), one could likewise optimize $u$ to maximize ${\bnorm{\tilde{A}}^2}$. However, the dependance of $\tilde{A}$ on disentanglers $u$ is not in a form that is directly compatible with the optimization problem discussed in Sect.\ref{sect:Project}. To this end, we use a modified environment $\tilde{\Gamma}_u$ of $u$ generated from ${\bnorm{\tilde{A}}^2}$, as depicted in Fig.\ref{fig:BigEnv}(c), where the modified environments results from the removal of either a pair tensors, $v_L v_L^{\dag}$ or $v_R v_R^{\dag}$, from $\tilde A$. The modified environment $\tilde{\Gamma}_u$ is now compatible with the previous optimization strategy of iterative SVD updates, thus can be used to directly replace the previous environment $\Gamma_u$ of $u$ from Fig.\ref{fig:Optimization}(c). Note that $\tilde{\Gamma}_u$ can be computed with cost that scales as $O(\chi^7)$ in terms of bond dimension $\chi$ [or $O(\chi^6)$ when employing the ideas of Appendix \ref{app:Reduce}], which is the same scaling with $\chi$ as the basic TNR algorithm discussed in Sect.\ref{sect:TNR}. 

\section{Achieving a scale-invariant RG flow} \label{sect:Scale}
A novel and useful feature of the TNR approach is that, when applied to a scale-invariant critical system, it can generate an explicitly scale-invariant RG flow, such that tensors at different scales $s$ of coarse-graining are equal,
\begin{equation} \label{eq:Ascale}
A^{(s+1)} \approx A^{(s)},
\end{equation}
up to small differences stemming from truncation errors. However, realization of scale invariance requires fixing the gauge degree of freedom, which we now discuss.

Give a square lattice network $\G$ composed of identical tensors $A_{ijkl}$ there exists a local gauge freedom in the network relating to a local change of basis on individual indices of the tensor, implemented by unitary matrices $\mathbf{x}$ and $\mathbf{y}$,
\begin{equation} \label{eq:Agauge}
A_{ijkl} \rightarrow \sum_{i'j'k'l'} (A)_{i'j'k'l'} \mathbf{x}_{ii'}(\mathbf{x}^{\dag})_{k'k} \mathbf{y}_{ll'}(\mathbf{y}^{\dag})_{j'j},
\end{equation}
under which the tensor network remains unchanged. [Note, in general the tensor network is invariant under changes of gauge implemented by invertible matrices, however the more restrictive class of unitary changes of gauge is sufficient to consider if reflection symmetry is exploited, see Appendix \ref{app:Reflect}]. If the gauge degree of freedom is not given proper consideration, application of TNR to a scale-invariant critical will not system will not, in general, produce an \emph{explicitly} scale-invariant RG flow, as defined by Eq.\ref{eq:Ascale}. Instead an \emph{implicitly} scale-invariant RG flow may be given, where the tensors $A^{(s+1)}$ and $A^{(s)}$ differ by choice of gauge, even though they are representative of the same critical fixed point. [Conversely, it can be demonstrated that TRG does not generate an implicitly scale-invariant RG flow when applied to a scale-invariant system, as certain gauge-invariant properties of tensors $A^{(s)}$ diverge with RG step $s$, see Ref.\onlinecite{TNR}]. We now explain how the gauge freedom in TNR can be fixed such that an otherwise implicitly scale-invariant RG flow becomes explicitly scale-invariant.

There are many strategies one could employ to fix the choice of gauge on tensor $A^{(s+1)}$ to be compatible with that on the previous tensor $A^{(s)}$. One possibility is to include separate gauge-fixing step after each coarse-graining iteration that minimizes the difference $\norm{A^{(s+1)} - A^{(s)}}$ through optimization of unitary matrices $\mathbf{x}$ and $\mathbf{y}$ that implement a change of gauge on $A^{(s+1)}$, as per Eq.\ref{eq:Agauge}. A different possibility, one that we find more convenient, is to include the gauge-fixing as part of the tensor optimization described in Sect.\ref{sect:Opt}. Let $B^{(s)}$ be the tensor obtained after the first step of the TNR iteration on $A^{(s)}$, as per Fig.\ref{fig:TNRsteps}(b), and assume we wish to choose a gauge on isometries $v_L^{(s)}$, $v_R^{(s)}$ and disentangler $u^{(s)}$ consistent with the previous TNR iteration, i.e. such that the choice of gauge on $B^{(s)}$ takes it as close to possible $B^{(s-1)}$. Let us define,
\begin{equation} \label{eq:Bfix}
\tilde B = B^{(s)} + \delta B^{(s-1)},
\end{equation} 
for some $\delta > 0$. Then, during the optimization of $v_L^{(s)}$, $v_R^{(s)}$ and $u^{(s)}$, if $\tilde B$ is used instead of $B^{(s)}$ in the calculation of the tensor environments, Fig.\ref{fig:Optimization}(a-c), the optimization is biased towards ensuring that the tensors are chosen in the same gauge as those at the previous TNR iteration. Typically we take $\delta = 1$ for the early stages of the optimization of tensors $v_L^{(s)}$, $v_R^{(s)}$ and $u^{(s)}$, but reduce $\delta$ smaller as the tensors converge. The same strategy can then be employed during the optimization of tensors $y_L^{(s)}$, $y_R^{(s)}$ and $w^{(s)}$ in the other intermediate steps of the TNR iteration to ensure that the gauge on $A^{(s+1)}$ is fixed in a way compatible with $A^{(s)}$.

\end{document}